\documentclass[acmsmall,authorversion,nonacm,screen,review=false,timestamp=false]{acmart}

\usepackage{amsmath, amsthm}
\usepackage{listings}
\usepackage{colortbl}
\usepackage{multicol}
\usepackage[framemethod=tikz]{mdframed}
\usepackage{dsfont}
\usepackage{caption}
\usepackage{multirow}

\usepackage[]{mdframed}
\newmdenv[%
    outerlinewidth  = .5,%
    roundcorner     = 5pt,%
    leftmargin      = 0,%
    rightmargin     = 0,%
    backgroundcolor = gray!20,%
    innertopmargin  = 5,%
    innerbottommargin = 5,%
    innerleftmargin   = 5,%
    innerrightmargin  = 5,%
    skipabove         = 5,%
    skipbelow         = 5%
]{rqanswer}

\newcommand{\infAdequate}{\mathit{infAdequate}}
\newcommand{\infer}{\mathit{infer}}

\newcommand{\randomTests}{\mathit{randomTests}}

\AtBeginDocument{%
  \providecommand\BibTeX{{%
    \normalfont B\kern-0.5em{\scshape i\kern-0.25em b}\kern-0.8em\TeX}}}

\setcopyright{rightsretained}
\acmDOI{10.1145/3660819}
\acmYear{2024}
\copyrightyear{2024}
\acmSubmissionID{fse24main-p1098-p}
\acmJournal{PACMSE}
\acmVolume{1}
\acmNumber{FSE}
\acmArticle{112}
\acmMonth{7}
\received{2023-09-28}
\received[accepted]{2024-04-16}

\definecolor{codegreen}{rgb}{0,0.6,0}
\definecolor{codegray}{rgb}{0.5,0.5,0.5}
\definecolor{codepurple}{rgb}{0.58,0,0.82}
\definecolor{backcolour}{rgb}{0.95,0.95,0.92}

\lstdefinestyle{mystyle}{
    backgroundcolor=\color{backcolour},   
    commentstyle=\color{codegreen},
    keywordstyle=\color{magenta},
    numberstyle=\tiny\color{codegray},
    stringstyle=\color{codepurple},
    basicstyle=\ttfamily\footnotesize,
    breakatwhitespace=false,         
    breaklines=true,                 
    captionpos=b,                    
    keepspaces=true,                 
    numbers=left,                    
    numbersep=5pt,                  
    showspaces=false,                
    showstringspaces=false,
    showtabs=false,                  
    tabsize=2
}

\begin{document}

\title{Bounding Random Test Set Size with Computational Learning Theory}

\author{Neil Walkinshaw}
\email{n.walkinshaw@sheffield.ac.uk}
\orcid{0000-0003-2134-6548}
\author{Michael Foster}
\email{m.foster@sheffield.ac.uk}
\orcid{0000-0001-8233-9873}
\author{Jos\'e Miguel Rojas}
\email{j.rojas@sheffield.ac.uk}
\orcid{0000-0002-0079-5355}
\author{Robert M. Hierons}
\email{r.hierons@sheffield.ac.uk}
\orcid{0000-0002-4771-1446}
\affiliation{%
\department{Department of Computer Science}
  \institution{The University of Sheffield}
  \streetaddress{Regent Court, 211 Portobello Street}
  \city{Sheffield}
  \country{UK}
  \postcode{S1 4DP}
}


\begin{abstract}
Random testing approaches work by generating inputs at random, or by selecting inputs randomly from some pre-defined operational profile. One long-standing question that arises in this and other testing contexts is as follows: When can we stop testing? At what point can we be certain that executing further tests in this manner will not explore previously untested (and potentially buggy) software behaviors? This is analogous to the question in Machine Learning, of how many training examples are required in order to infer an accurate model. In this paper we show how probabilistic approaches to answer this question in Machine Learning (arising from Computational Learning Theory) can be applied in our testing context, to provide an upper-bound on the number of tests required to achieve a given level of adequacy. We validate this bound on a large set of Java units, and an autonomous driving system.
\end{abstract}



\maketitle

\section{Introduction}

Random testing \cite{hamlet1994} is a widely used approach to automated software testing. It relies on the availability of an efficient process for generating test inputs that can be readily used to generate and execute large numbers of tests \cite{arcuri2011random}. The mechanisms that are used for this can range from entirely uniform and genuinely random, to more carefully crafted generators that produce inputs that collectively represent an operational profile of the program under test \cite{musa1993operational}. 

One fundamental question that arises with these automated approaches is \cite{goodenough1975toward,frankl1988applicable}: When can the testing be stopped? When do we hit a `saturation point' after which the execution of further tests will not cover any additional functionalities or potentially expose additional bugs? Currently, the decision of when to stop is often left to the judgment of the individual tester \cite{bohme2019assurances}, or testing is afforded a fixed time-budget \cite{tramontana2019developing}. In their roadmap papers, Harrold \cite{harrold2000testing} and B\"ohme \cite{bohme2019assurances} make the case for frameworks that provide a statistical basis for test adequacy and stopping criteria.


A similar question arises in the context of Machine Learning \cite{valiant1984theory,valiant2014probably}: How many data-points are required to reliably infer an accurate model? In this context, Computational Learning Theory enables us to place bounds on the number of training examples that would be required by a given learning algorithm to guarantee some level of accuracy. Valiant's Probably Approximately Correct (PAC) framework \cite{valiant1984theory} has been used to establish bounds for several learning algorithms.

These PAC bounds are relevant because this relationship between testing and Machine Learning was established 40 years ago, almost simultaneously by Weyuker \cite{weyuker1983assessing} and Budd and Angluin \cite{budd1982two}. They suggested that the two fields could be combined; that test-adequacy could be assessed by inferring a predictive model of software behavior from the observed test inputs and outputs. If the model in question is accurate, then the test set clearly has a sufficiently high information content, and can be deemed to be  ``inference adequate'' \cite{weyuker1983assessing,zhu1996formal}. In this context, the requirements on the number of examples for Machine Learners established by PAC reasoning can potentially be used to establish the number of tests that would be required to reach a point of saturation.


There have been sporadic efforts to apply PAC to testing, most of which took place in the 90s \cite{zhu1992inductive,zhu1996formal,romanik1996using}. These  are however hard to apply to generic software. They rely on the availability of a learning algorithm that is both sufficiently versatile to infer the input-output behavior for arbitrary systems (a task that remains elusive in and of itself), \emph{and} has been proven to be a PAC algorithm -- i.e. to be able to guarantee an accurate result within a polynomial number of training examples.

In this paper we show how the above problem can be side-stepped. Instead of characterizing the tested behavior of a program in terms of externally observable behaviors, we encode the behavior as the coverage of test objectives within the source code. The choice of objectives is flexible, and may comprise source code statements, branches, etc. The salient point for us is that these can be coded as simple lists of  Boolean variables (indicating whether an objective has / has not been covered), regardless of the externally observable behavior of the underlying program. 
 
 In this context, we show how the question of how many tests are required before a point of saturation is reached can (in the context of random testing) be framed as the question of how many training examples would be required to infer a Boolean conjunction, where the Boolean variables in question correspond to coverage objectives. Since Boolean conjunctions have been shown to be PAC-learnable \cite{valiant1984theory}, we show how the upper-bounds on the training sets for Boolean conjunctions can be mapped to upper bounds on the number of tests required to achieve an adequate test set. 

In effect, this enables us to calculate the number of tests that must be executed in order to offer a probabilistic \emph{guarantee} that the underlying test set is adequate. This is done `\emph{a-priori}' -- without the need to execute any tests or to run Machine Learning algorithms. The specific contributions of this paper are as follows:

\begin{itemize}
    \item We show how the question of establishing which statements are executable by a random test set \cite{miranda2020testing} maps to the Machine Learning problem of inferring Boolean conjunctions \cite{valiant1984theory}.
    \item We show how PAC limits that bound the number of training examples required to guarantee accuracy of inferred Boolean conjunctions can be applied to bound the number of random tests beyond which there is a saturation effect.
    \item We empirically demonstrate the reliability of the bound in two studies: One on random tests of 7,198 publicly available Java units, and another on operational tests of an autonomous driver in the CARLA driving simulator.
\end{itemize}

\section{Testing and Test Adequacy}

\subsection{Random Testing}
\label{sub:random}

Random testing  \cite{hamlet1994} is a popular approach to test-input generation. It is based on the premise that inputs to a program can be characterized in the form of a distribution, and that inputs can be repeatedly sampled.  Ideally, the distribution from which tests are selected should reflect the intended `operational profile' of the system \cite{musa1993operational}. This makes it possible to draw probabilistic conclusions about the reliability of the system once it is deployed (i.e. the likelihood of failures arising, assuming that the operational profile is accurate \cite{hamlet1994}) from the sampled test executions.

\begin{definition}
\label{def:randomDistribution}
We define a random test generator as the function\footnote{We assume that this function is non-deterministic, and that the sample varies each time. We choose this notation because it is aligned with equivalent data-sampling definitions used in PAC, which are covered later.} $\randomTests:I \times D_I \times n \rightarrow T$. $I$ represents the input domain - the set of possible inputs to the system under test. $D_I$ represents a distribution over these elements. Since we will encounter distributions of elements in a Machine Learning context as well, we will use the subscript (e.g. $I$) to signify the set of elements to which the distribution applies. $n$ represents the number of times that $D_I$ is sampled with a uniform probability (also known as `sampling with replacement'). The result is a list of test inputs $T=[t_1, \ldots, t_n]$.
\end{definition}


\subsection{Test Adequacy and Code Coverage}
\label{sub:adequacy}

Test data adequacy \cite{goodenough1975toward} refers to a measure or a criterion to evaluate the capability of a test set to reliably expose the faults in a system.
If a test set is deemed to be adequate, a failure-free execution of the test set should ideally imply that the system under test does not contain any faults. 

\subsubsection{Code Coverage}

The most popular approach to gauging test adequacy is to measure code coverage. This is founded on the premise that there is a relationship between the extent to which the source code is executed, and the capability of a test set to explore and expose faults. Many code-based test adequacy measures have been proposed over the years \cite{zhu1997software}. In practice, statement and branch coverage remain the most commonly used \cite{ivankovic2019code}.

Though widely used, code coverage has been the subject of a significant amount of criticism. Several studies have indicated a relatively poor correlation between code coverage and the ability of a test set to expose faults \cite{inozemtseva2014coverage,chekam2017empirical}. One key criticism is that it is not decidable whether all of the code elements are executable in the first place \cite{frankl1988applicable,budd1982two}. This gives rise to a dilemma for the tester \cite{weyuker1986axiomatizing}: if a test set achieves only 80\% coverage, is this because the remaining 20\% are not feasibly executable? Or is it because the test set is incomplete? 

\subsubsection{Predicting Test Saturation}
\label{subsub:saturation}

One well-established property of random testing is known as the \emph{saturation effect} \cite{lyu1996handbook,tramontana2019developing}. As the number of tests increases, the rate of convergence towards a given adequacy criterion decreases. Beyond a certain point, any change in coverage is only negligible. In other words, when running random tests, there is invariably a point of `diminishing returns', where executing further tests to improve coverage is no longer worthwhile. 

Tracking coverage can play a useful role in assessing the point at which a saturation point has been reached \cite{amalfitano2015exploiting}. Depending on the type of system and testing approach, this can also take a long time. In a recent experiment by Liyanage \emph{et al.} \cite{liyanage2023reachable}, no saturation point was observed over the span of a week long fuzzing cycle. 

The question then arises of whether the saturation point can be predicted, or a bound can be placed on the number of tests required to reach the saturation point. Arcuri and Briand \cite{arcuri2011random} re-framed the task of meeting test-objectives as the Coupon Collector problem \cite{motwani1996randomized}, in order to place a lower bound on the number of tests required.  However, their Coupon-Collector bound has two significant drawbacks. Firstly, it is dependent upon the assumption that test objectives are disjoint; i.e. that a given test will only achieve a single test objective if any. As such, scenarios where a single test can achieve multiple objectives (e.g. covering multiple statements or branches) are not applicable. Secondly, it is only a \emph{lower} bound: the number of test cases that are actually required to meet the test objectives could be significantly higher. In practice, a tester would ideally have an \emph{upper} bound, i.e., a number of tests beyond which they can guarantee (perhaps within some given probability) that their testing objectives have been met.

There has also been a significant amount of work on predicting saturation in the context of fuzzing, by Liyanage \emph{et al.} \cite{liyanage2023reachable,liyanage2024extrapolating}.  In this context, Liyanage \emph{et al.} proposed an `on-line' approach that can predict the saturation point on the basis of coverage data from tests executed so far. They show that the predictions converge on an accurate estimate after an initial `burn-in' period. Their fuzzing context is different to the standard random-testing scenario considered here, because fuzzers tend to actively adapt their seed schedules to maximize coverage \cite{zhu2022fuzzing}. Since they are dealing with an input-distribution that is dynamic, the relationship to code coverage requires a significant amount of data to estimate (this is corroborated by their results, which can require millions of executions before they can converge on an accurate prediction of saturation). 


\subsubsection{Relevant Coverage and its Relation to Saturation}
\label{subsub:relevant}

In practice, testers are only assessing test effectiveness in terms of code that is feasibly executable. In the random testing context (Definition \ref{def:randomDistribution}), this amounts to the  sub-set of statements that can be executed by inputs in $I$ that have a non-negligible probability of being selected according to $D_I$. The notion corresponding to this sub-set of statements is referred to by Miranda and Bertolino \cite{miranda2015improving,miranda2020testing} as  `Relevant Coverage'. 

Relevant Coverage is directly related to the saturation effect discussed above. The achievement of Relevant Coverage coincides with the point at which the saturation point has been reached. As such, the ability to predict or bound the number of tests required to achieve Relevant Coverage implies the ability to predict the number of tests at which saturation occurs.



\subsubsection{Program Spectra}

Program spectra \cite{reps1997use,harrold1998empirical} offer a more fine-grained perspective on software behavior. A program spectrum relates individual executions to code elements. 
The data recorded for a code element typically includes  whether a code element has been executed (a Hit Spectrum), or the number of times it has been encountered during a single execution (known as a Frequency or Count Spectrum). Program spectra have proven to be particularly useful for fault localization \cite{wong2016survey}. 

In their work on Relevant Coverage, Miranda and Bertolino show that a Hit Spectrum can be used as a basis for determining the set of statements that should be `in-scope' in order to measure the relevant coverage \cite{miranda2020testing}. The premise is that, for a given set of test cases, a hit spectrum can be used to distinguish between statements that are or are not relevant to a usage context. If a statement is hit by every test case for the usage-context in question, it can be deemed to be essential and in-scope. Otherwise, if it is only occasionally executed, it can be deemed to be out of scope. 

In this paper we take advantage of this relationship between hit-spectra and coverage. To facilitate this discussion later on, we provide a more formal definition of hit spectra here.

\begin{definition}
\label{def:hitspectrum}
For a program with a set of $n$ code elements (we will be referring to statements by default), which has been executed by $t$ test cases, a Hit Spectrum $HS$ is a binary matrix: 
$$HS = 
\begin{bmatrix}
x_{1,1} & \ldots & x_{n,1}\\
\vdots & \ddots & \vdots\\
x_{1,t}& \ldots & x_{n,t}
\end{bmatrix}
$$
An element $x_{i,j}=1$ if test execution $i$ executes code element $j$, and $x_{i,j}=0$ otherwise. 

\end{definition}

\subsection{Inference Adequacy}
\label{sub:inferenceAdequacy}

The link between testing and Machine Learning can be traced back to Moore's 1956 paper on Gedanken Experiments on Sequential Machines \cite{moore1956gedanken}. The relationship was however first formalized 40 years ago by Weyuker \cite{weyuker1983assessing}, who posited (along with Budd and Angluin \cite{budd1982two}) that inference could be used for assessing test adequacy. We provide a version of Weyuker's formalization here (customized to facilitate the linkage with Probably Approximately Correct learning in Section \ref{sec:learningTheoryTesting}).

The essential idea behind inference adequacy is that a test set can be deemed to be `adequate' if, upon execution, it generates sufficient information to enable an inference engine to infer a model that accurately represents the system. To formalise this we start by defining the system under test (SUT). This has an input domain $I$ (as defined for random test generation in Definition \ref{def:randomDistribution}). We use $O$ to denote the set of observable outputs.  

\begin{definition}
\label{def:sut}
The SUT  is defined as a function  $sut: I \rightarrow O$.
\end{definition}

\begin{definition}
\label{def:observationTrace}
An \emph{observation trace} $Obs$ is obtained by running a list of inputs $t \in I$ on the $sut$. $Obs=[(t_1,sut(t_1)),\ldots,(t_n,sut(t_n))]$, where $n=|T|$.
\end{definition}

\begin{definition}
\label{def:infer}
Given a set of test observations $Obs$, an inference engine $\infer: Obs \rightarrow M$ will infer a model $M:I \rightarrow O$, which is capable of predicting outputs in $O$ for inputs in $I$.
\end{definition}

\begin{definition}
\label{def:inferenceAdequacy}
\emph{`Inference adequacy'} can be defined as a Boolean valued function, as follows\footnote{In her paper Weyuker actually provides two flavors of inference adequacy - one where the inference involves the provision of a specification stipulating correct behavior properties to be inferred, and one where the behavior is evaluated directly in comparison with the program. We focus here on the latter as we do not in this paper presume the provision of a specification.}:

$$\infAdequate: T \times sut \times \infer \rightarrow \mathds{B}$$

$T$ represents the set of test inputs that we seek to assess. $sut$ and $\infer$ are as presented in Definitions \ref{def:sut} and \ref{def:infer} respectively. The tests in $T$ are run on the SUT to produce a list of observations, which is used to infer model $M$. $\infAdequate$ returns \texttt{true} if $M \equiv sut$, and \texttt{false} otherwise. 
\end{definition}

Establishing this equivalence  $M \equiv sut$ is challenging in practice, because it is essentially a model-based testing problem in and of itself. One typical approach (also proposed by Weyuker) is to create a further `reference' test set $R$ (Weyuker suggests that this may be done by random generation \cite{weyuker1983assessing}). $\infAdequate(T,sut,\infer) = \texttt{true}$ iff $\forall r \in R, sut(r) = M(r)$. 

This view of test adequacy is distinctive because it is not coverage-based. The task is to collect enough information to enable the inference of a model $M$ such that $M \equiv sut$. This circumvents the issues arising with code coverage mentioned in Section \ref{sub:adequacy}.

There has been a significant amount of research into testing techniques that apply this principle \cite{bergadano1996testing,ghani2008strengthening,briand2009using,meinke2011incremental,choi2013guided,isberner2015open,walkinshaw2017uncertainty}.
Typically, they adopt an iterative approach of test-selection, execution and inference. The role of $sut$ is usually played by the program under test itself, and no presumptions are made about the provision of specification documents (though there are exceptions to this \cite{dupont2008qsm}). The inferred model $M$ is evaluated in terms of its predictive accuracy (either by further tests, or by using a Machine Learning evaluation process such as k-folds cross-validation). If the model is deemed to be sufficiently accurate, the process terminates. The key point is that the test-inference loop depends on the output of an inference algorithm.

There are three strong caveats that apply when using inference adequacy. Firstly, this form of adequacy is only \emph{internally} valid. The accuracy of the model (and therefore the inference-adequacy of the test set) can only be assessed in terms of $R$, the second evaluative test set. This however leads to a circular situation where there is no reason that we can have any more confidence in the adequacy of $R$ than we can in $T$.

Another more technical restriction of this approach is the fact that it is limited to programs where there exists an appropriate inference algorithm for the observable behavior. For many classes of program the choice of inference algorithm is not obvious. Currently, the vast majority of approaches tend to focus on programs where the behavior can be captured by reasonably straightforward classes of models such as finite state machines or decision trees. 

A final consideration is the trade-off between the time required to infer a model, and the respective benefit of using this time to simply execute larger numbers of tests instead. This point was raised by Arcuri and Briand \cite{arcuri2011random} with respect to Adaptive Random Testing \cite{chen2005adaptive}. The value of inference adequacy will be more pronounced for programs with longer execution times, where the time required to infer and reason about a model is more worthwhile than simply using the inference time to execute large numbers of random tests.

\section{Computational Learning Theory}

Computational Learning Theory refers broadly to the analysis of the theoretical capabilities of Machine Learning algorithms. This involves the analysis of whether a given concept is learnable in theory and, by implication, whether there is an asymptotic bound on the number of data-points required by a learning algorithm to achieve this. Valiant's Probably Approximately Correct (PAC) framework \cite{valiant1984theory,valiant2014probably} offers a particularly useful and well-established basis for such analysis.

We will provide a brief summary of PAC, and the  reasoning involved in establishing a bound on the number of examples required by a learner. We base our presentation of PAC on the notation used by Shalev-Shwartz and Ben-David in their book on Machine Learning foundations \cite{shalev2014understanding}.

\subsection{The Probably Approximately Correct Framework}
\label{sub:pac}

The following definitions capture the PAC learning setting (as discussed in Learning theory text books \cite{kearns1994introduction,shalev2014understanding}) in such a way that will facilitate the linkage with testing in later sections.

We start from a \emph{domain set} of objects $X$ which will form potential inputs to an inferred model (e.g. bitmaps of hand-written characters for which we wish to infer the character). For training purposes, we collect a sample of these according to some fixed distribution $D_X$, within which elements are independently and identically distributed (i.i.d.).

\begin{definition}
\label{def:sample}
    $sample: X \times D_X \times m \rightarrow S$ represents the process of sampling a set $S$ of $m$ elements in $X$ according to $D_X$.
\end{definition}

It is also assumed that there is a set of labels $Y$, representing the set of possible labels that could be attributed to elements in $X$ (e.g. the set of alphanumeric characters). 

\begin{definition}
\label{def:label}
    $label: X \rightarrow Y$ represents the process of attributing the `ground truth' label in $Y$ for some given element in $X$.
\end{definition}

A learner's objective is to infer a model that, for any $x$ in $X$, accurately predicts the label in $Y$. 

\begin{definition}
    \label{def:trainingSet}
    A \emph{training set} is constructed by the function $tr$, which takes a sample of inputs with the $sample$ function, and identifies the corresponding labels for them with the $label$.   $tr: sample \times label \rightarrow [(x_1,label(x_1), \ldots, (x_m, label(x_m))]$, where $x_n \in X$ and $label(x_n) \in Y$. 
\end{definition}

\begin{definition}
\label{def:hypothesisClass}
The hypothesis class $H$ for a learner corresponds to the set of all possible mappings $X \rightarrow Y$ representing potential target concepts in the instance space.  
\end{definition}

\begin{definition}
\label{def:learner}
 A learner $L: TS \rightarrow H$ denotes the learner. This takes as input a training set $TS$ and returns a hypothesis $h \in H$.
\end{definition}

\begin{definition}
\label{def:generalization}

Ideally an inferred hypothesis should have a small \emph{generalization error}. The generalization error $Loss_{D,label}(h)$ is the probability that $h$ does not predict the correct label on a random data point generated by the underlying distribution $D$. 

\begin{equation}
Loss_{D,label}(h) = \mathds{P}_{x \sim D}[h(x) \neq label(x)] = D(\{x:h(x) \neq label(x)\})
\end{equation}
\end{definition}

Since  distribution $D$ and labeling function $label$ are not known to the learner, the generalization error can only be approximated by calculating the \emph{empirical error} with respect to labeled sample $S$.

\begin{definition}
    The \emph{empirical error} is the proportion of instances in $S$ that are misclassified:

    \begin{equation}
    Loss_S(h) = \frac{|\{i \in [1..m]:h(x_i) \neq y_i\}|}{n}
    \end{equation}
\end{definition}



For a concept class to be ``learnable'' in a PAC setting, the objective is to prove that an algorithm will always return a result where the accuracy is reasonably high. For this we wish to ensure that the generalization error (Definition \ref{def:generalization}) is bounded by some allowable margin of error $\epsilon$. Since the samples in the training set are sampled i.i.d. from some distribution D, there is also a chance that a sample is chosen that is not representative of the target concept $c$. In a PAC setting, we want to ensure that the risk of this falls within a bound $\delta$. So the \emph{confidence} that we are inferring from a representative distribution is denoted $(1-\delta)$.

In a PAC setting, it is normal to adopt the simplifying assumption that every hypothesis returned by a learning algorithm in PAC is \emph{consistent} \cite{shalev2014understanding}: it is entirely consistent with the training sample $S$ ($L_S(h)=0$). This assumption (referred to as the Realizability assumption) can be relaxed \cite{shalev2014understanding}, but is kept here because it simplifies the discussion of the sample complexity.

\begin{definition}
\label{def:pac}
    A hypothesis class $H$ is said to be PAC-learnable if there exists a `sample complexity' function $m_H:(\epsilon,\delta)$ (discussed in more detail below) and a learning algorithm with the following property: For every $\epsilon, \delta \in (0,1)$, for every distribution $D$ over $X$, and for every labelling function $label:X \rightarrow \{0,1\}$, if the realizability assumption holds with respect to $H, D, label$, then when running the learning algorithm on $m \geq m_H(\epsilon,\delta)$ i.i.d. examples generated by $D$ and labelled by $label$, the algorithm returns a hypothesis $h$ such that, with probability at least $1-\delta$ (over the choice of examples), $Loss_{D,f}(h) \leq \epsilon$.
    
\end{definition}

\subsection{Hypothesis Space Size and Sample Complexity}
\label{sub:hypothesisSpace}

The function $m_H:(0,1)^2 \rightarrow \mathds{N}$ determines the \emph{sample complexity} of learning $H$. It determines the number of examples (i.e. the size of $S$) that  \emph{guarantee} a probably approximately correct solution \cite{shalev2014understanding}. The ability to prove that some concept is PAC-learnable implies the ability to derive a bound for $m_H$. This depends on the choices for $\epsilon$ and $\delta$, and an ability to quantify the complexity of $H$.

When it comes to quantifying the complexity of $H$, in the simplest case (which is sufficient for the purposes of our paper), we consider the situation where $H$ is finite. The size of $H$ may be a (potentially exponential) function of $m$ - the computational cost of representing an instance $x \in X$, e.g. the number of features in an instance. We will provide an example of such a bound below. 

When $H$ is finite, $m_H$ can be bounded as follows \cite{haussler1988quantifying}:

\begin{equation}
\label{eq:bound}
m \geq \frac{1}{\epsilon} \left( ln(|H|) + ln \frac{1}{\delta} \right) \\
\end{equation}

In other words, if $m$ satisfies this bound, then we can guarantee that, for any inferred hypothesis $h$, the generalization error will be less than or equal to $\epsilon$, with a probability of $1-\delta$. It is important to emphasize that this pertains to the generalization error as opposed to the empirical error; we can be confident that any model inferred from a sample of size $m$ will be probably approximately accurate without the need to sample data or execute a learner. 

The proof of this limit is detailed by Haussler \cite{haussler1988quantifying} and in text books on learning theory \cite{kearns1994introduction,shalev2014understanding,mohri2018foundations}. These references also contain generalizations of this bound that apply to situations where $H$ is infinite (which draw upon the Vapnik Chervonenkis dimension \cite{vapnik1999nature} as a means of measuring the complexity of the hypothesis space).

\subsection{Boolean Conjunctions}
\label{subsub:conjunctions}

We focus on the case of finite hypothesis spaces, because this is all that we require. To illustrate the concept of a hypothesis space and the corresponding sample complexity, we consider the specific challenge of learning Boolean conjunctions. Aside from being one of the canonical examples of PAC learnability \cite{valiant1984theory,kearns1994introduction}, it is also the algorithm that we use to reason about test set size.

For the conjunction learning setting, the instance space is $X_n=\{0,1\}^n$. Each training instance $a \in X_n$ is an assignment to the $n$ boolean variables $x_1, \ldots, x_n$. An example instance might be the conjunction $x_1 \wedge \neg x_3 \wedge x_4$ \cite{kearns1994introduction}, which would be represented as the set $\{a \in \{0,1\}^n:a_1=1, a_3=0, a_4 = 1\}$. Note that this leaves the possibility that some variables will not be given any assignments. The task for a learner would be to, from a set of instances in $X$, produce a hypothesis $h$ where $h$ is some specific target conjunction. In this case, the size of the hypothesis space (denoted $|H_n|$)  is $|H_n|=3^n$ -- where each variable is either \texttt{true}, \texttt{false}, or is not given any assignments. 

An algorithm that renders $H$ PAC-learnable \cite{valiant1984theory} could start from a hypothesis that is the conjunction of all positive and negative literals (i.e. $h=x_1 \wedge \neg x_1 \wedge \ldots \wedge x_n \wedge \neg x_n$). Then, for each example from $D$, $L$ updates $h$ such that any literal that `contradicts' the example is deleted: for each $i$, if $a_i=0$, $x_i$ is deleted and if $x_i=1$, $\neg x_i$ is deleted. The proof that this renders the boolean conjunction problem PAC-learnable is elaborated by Valiant \cite{valiant1984theory,valiant2014probably} and Kearns and Vazirani \cite{kearns1994introduction}.

In order to establish the  number $m$ of examples  that would be required to \emph{guarantee} an accurate (within error bounds $\epsilon$ and $\delta$) conjunction, we can therefore apply the limit from Equation \ref{eq:bound}. For example, for an expression of 30 boolean literals ($n=30$), with $\epsilon=0.1$ and $\delta=0.1$ (we tolerate a classification error of up to 10\%, and require this to be upheld 90\% of the time), this would yield\footnote{For this specific learning task there is a slightly tighter bound defined by Kearns and Vazirani \cite{kearns1994introduction}, but we opt for this more generally applicable limit for the purposes of this paper.}:

$$m \geq \frac{1}{0.1}(ln(30^3) + ln(1/0.1))=352.6095$$

\section{Applying Computational Learning Learning Theory to Testing}
\label{sec:learningTheoryTesting}

The ability within Computational Learning Theory to place an upper bound on the number of data-points required to guarantee an accurate model is highly pertinent from our random testing perspective. Given the established relationship between Testing and ML (Section \ref{sub:inferenceAdequacy}), this implies that it should be possible to apply the same bounds to the number of tests required to reliably test a software system. 



In the remainder of this section, we join these concepts together. We show how the PAC bounds that apply to Boolean conjunctions can be used to bound the number of random tests required to ensure (subject to PAC parameters) Relevant Coverage. We start in Section \ref{sub:inferenceAdequacyAndPAC} by formalizing the relationship between Inference Adequacy and PAC learning. In Section \ref{sub:pacCoverage} we show how Hit Spectra can serve as a useful, learnable encoding of test cases, and how these relate to the challenge of Boolean conjunction inference. This is followed by a discussion of some caveats and assumptions that apply to the approach in Section \ref{sub:caveats}, and finally by a worked example in Section \ref{sec:randomBounds}.

\subsection{Inference Adequacy for Random Testing as a PAC Learning Problem}
\label{sub:inferenceAdequacyAndPAC}

\newcommand{\specialcell}[2][c]{%
  \begin{tabular}[#1]{@{}c@{}}#2\end{tabular}}

\begin{table}
\caption{Key correspondences between Inference Adequacy and the PAC framework.}
\label{tab:pacTesting}
\begin{smaller}
    \begin{tabular}{l c c} \toprule
        &\textbf{Inference Adequacy} & \textbf{PAC} \\
        \midrule
        \textbf{Data} & \specialcell{$I$, $O$ \\Inputs and outputs} & \specialcell{$X$, $Y$\\Examples and labels}\\
        \midrule
        \textbf{Data source} & 
        & \specialcell{$sample: X \times D_X \times m \rightarrow S$\\ \emph{(Definition \ref{def:sample})}} \\
        \midrule 
        \textbf{Labeling procedure} & \specialcell{$sut:I \rightarrow O$ \\ (Definition \ref{def:sut})}& \specialcell{$label:X \rightarrow Y$ \\ (Definition \ref{def:label})}\\
        \midrule
        \textbf{Training set} & \specialcell{Observation trace $Obs$ \\ (Definition \ref{def:observationTrace})} & \specialcell{$tr: sample \times label \rightarrow TS$ \\ (Definition \ref{def:trainingSet})}\\
        \midrule
        \textbf{Inference} & \specialcell{$\infer:Obs \rightarrow M$ where \\
        $M: I \rightarrow O$\\(Definition \ref{def:infer})} & \specialcell{$L:TS \rightarrow H$ where \\ (Definition \ref{def:learner}) \\\\ $H$ is all possible  mappings  $X \rightarrow Y$ \\(Definition \ref{def:hypothesisClass})}  \\
        \bottomrule
    \end{tabular}
    \end{smaller}
\end{table}

The main correspondences between inference adequacy and PAC are summarized in Table \ref{tab:pacTesting}. There are two noteworthy distinctions between the two. Firstly, inference adequacy does not make any presumptions about the process used to generate the test set (the data source in Table \ref{tab:pacTesting}); $Obs$ can in principle be any test set. For PAC, training samples are selected i.i.d. from a fixed distribution. This distribution  enables the calculation of the sample-complexity (thanks to the ability to apply the Chernoff Bounds \cite{kearns1994introduction}). Secondly, the inference adequacy definition focuses on the inference of some concrete model $M$, whereas PAC is concerned with an entire hypothesis class of models $H$.

If we can guarantee that tests are sampled from a fixed distribution, then we can apply the PAC bounds in a testing context. This is straightforward for random testing: the random test generation definition $\randomTests:I \times D_I \times n \rightarrow T$  corresponds to the definition of $sample$ in the PAC context.

If we restrict attention to tests generated by $\randomTests$, then the training set used can be re-defined as a function, analogously to the definition of $tr$ (Definition \ref{def:trainingSet}):

\begin{definition}
    $tr_{rnd} : \randomTests \times sut \rightarrow TS$, where $TS$ is as defined in Definition \ref{def:trainingSet}.
\end{definition}

We can then interpret hypothesis class $H$ as the set of all possible mappings $I \rightarrow O$ (analogous to Definition \ref{def:hypothesisClass}). This, then, enables us to characterize random testing in terms of PAC-learnability. As per Definition \ref{def:pac}, if $H$ is PAC-learnable, it is possible to apply the sample-complexity function $m_H$ described in Section \ref{sub:hypothesisSpace}. In the testing context, this provides us with an upper-bound on the number of tests required to \emph{guarantee} inference-adequacy with respect to parameters $\epsilon$ and $\delta$.

To apply this limit in practice, we need to be able to determine the size of the hypothesis space $|H|$. However, as discussed in Section \ref{sub:inferenceAdequacy}, current inference-adequacy approaches are based on the inference of  models of software behavior and there is no one-size-fits-all approach to determining $|H|$. Furthermore, the range of hypothesis classes that have been shown to be PAC-learnable is very restricted. Even conceptually simple notions such as the potential sequencing of events as expressed by finite state machines are not PAC learnable \cite{gold1978complexity,kearns1994introduction}, unless significant additional assumptions are made about the learning context \cite{angluin1987learning}. Accordingly work on relating PAC to testing by Zhu \emph{et al.} \cite{zhu1996formal} and Romanik \emph{et al.} \cite{romanik1996using}, was restricted to fragments of programs or highly restricted types of programs.



\subsection{Hit Spectra as a Basis for Inference Adequacy}
\label{sub:pacCoverage}

We are specifically interested in reasoning about the number of tests that are required to reach a saturation point (see Section \ref{subsub:saturation}). Given that saturation is defined as the point at which there are no further increases in code coverage, we are particularly interested in capturing program behavior in such a way that relates test executions to coverage.

\subsubsection{Encoding Tested Behavior with Hit Spectra}

This is where Program Spectra (and specifically Hit Spectra) play an important role for us. Hit spectra explicitly relate test executions to statements that are (resp. are not) covered. They are very simple, can be generated automatically, and capture all of the information that would be required in practice to determine whether testing has reached a saturation point. The value of hit spectra for reasoning about tested behavior has been demonstrated in recent work by Bertolino \emph{et al.} \cite{bertolino2019adaptive}.


To encode tested behaviors in the form of a hit spectrum, we adopt Definition \ref{def:hitspectrum}. The resulting hit spectrum $HS$ is a binary matrix, where each row represents a single test execution and each column represents a code element. For each test execution (row), a column has a 1 if the element is covered by the test case, and a 0 otherwise. 

It is worth emphasizing that, using this abstraction of behavior, we do not need to consider inputs and outputs. A test execution can be conceptualized entirely in terms of the statements that it does / does not cover. Since we are in a setting where the test cases are selected at random from some distribution $D_I$, there is no need to reason about inputs.

\subsubsection{Framing Relevant Coverage as a Boolean Conjunction Learning Problem}

Our objective is to determine the number of tests that are required to reach a saturation point -- the point at which there is no change in code coverage. In terms of the Hit Spectrum, the saturation point happens at the point at which the addition of tests (rows in the Hit Spectrum) does not lead to any columns (representing code elements) being assigned a value of 1 if previously all of their values have been 0.

In this, there is a direct mapping to the Boolean Conjunction inference setting (Section \ref{subsub:conjunctions}).  If we consider Relevant Coverage (Section \ref{subsub:relevant}), the set of statements that are relevant for some input distribution $D_I$ are those that are executed by at least one of the test cases from $D_I$. Conversely, the statements that are not relevant is the set of statements that are never executed by any of the test cases. For a program with $n$ code elements in total, the subset of relevant statements can be expressed in the form of a conjunction of $n$ boolean variables $x_1, \ldots, x_n$. For example, for $n=3$, a conjunction $x_1 \wedge \neg x_3$ would indicate that statement 1 is executed by all of the tests, $x_2$ is not in the conjunction because it is covered by some but not all of the tests, and $x_3$ is not covered by any of the tests (and is therefore not considered a part of the Relevant Coverage).

\subsubsection{Computing a Bound on the Number of Tests}

We know that Boolean conjunctions are PAC-learnable \cite{valiant1984theory}, and that we are therefore able to derive a sample-complexity. In our testing context, we know the number of Boolean literals that might exist in our conjunction - any of the $n$ code elements in our program. This means that the hypothesis space $|H|=n^3$ (see Section \ref{subsub:conjunctions}). Accordingly, the upper bound on the number of test cases can then be obtained by Equation \ref{eq:bound}.



\subsection{Worked Example: The Triangle Classification Problem}\label{sec:randomBounds}
{
\begin{figure}
\begin{mdframed}[backgroundcolor=backcolour,
    hidealllines=true,%
    innerleftmargin=0.1cm,
    innerrightmargin=0.1cm,
    innertopmargin=-0.3cm,
    innerbottommargin=-0.1cm]
    \begin{multicols}{2}
    \begin{lstlisting}[language=Java]
int tri_type(int a, int b, int c) {
    int type = NOT_A_TRIANGLE;
    if (a > b) 
        int t = a; a = b; b = t; 
    if (a > c) 
        int t = a; a = c; c = t; 
    if (b > c)  
        int t = b; b = c; c = t; 
    if (a + b > c) 
        type = checkType(a,b,c);
    return type;
}

int checkType(int a, int b, int c) {
    int type;
    if (a + b <= c) 
        type = NOT_A_TRIANGLE;
    else {
        type = SCALENE;
        if (a == b && b == c) 
            type = EQUILATERAL;
        else if (a == b || b == c) 
            type = ISOSCELES;
    }
    return type;
}
\end{lstlisting}
\end{multicols}
\end{mdframed}
\caption{Triangle classification source code.\label{lst:triangle}}
\end{figure}
}

To illustrate the approach, we will use the classical triangle classification example. The version we use (Listing \ref{lst:triangle}) includes an unreachable statement: Line 17 (setting the return variable to indicate that the triangle is not valid) in the \texttt{checkType} function can never be executed because the predicate at line 9 in the \texttt{tri\_type} client function only allows the call to \texttt{checkType} to occur if the triangle is valid. Using conventional statement coverage, the goal of 100\% coverage would be unachievable. If we wanted to apply Miranda and Bertolino's `relevant coverage' \cite{miranda2020testing} (Section \ref{sub:adequacy})\footnote{Miranda and Bertolino actually make a distinction between statements that are unreachable, and statement that are irrelevant to $D_I$, and explicitly focus on the latter. In our case we consider both; we are only interested in statements that are relevant to the usage context, and executable.}, we would need some way of determining that line 17 should not be `in scope'.

\paragraph{Test execution representation}

Each test is represented as a list of $n$ Boolean literals $x_1, \ldots, x_n$, where $n$ corresponds to the number of test objectives in the program under test. For a given test, any statement $x_i$ that is not executed is assigned \texttt{false}, and any statement that is executed is \texttt{true}.

\paragraph{Applying the Boolean Conjunction inference algorithm to Coverage Spectra}

We sample our tests from a uniform distribution over parameters a, b and c,  where each parameter is in the range 1 to 10\footnote{Our choice of range is somewhat arbitrary for the purposes of this illustration, and the nuances involved in setting the parameters for an input distribution are discussed elsewhere \cite{hamlet1994,arcuri2011random}}. The full generated test set along with coverage results is available online\footnote{\url{https://docs.google.com/spreadsheets/d/e/2PACX-1vRa_xuroS6AJIWNLPFD3gBPxjUZv-5b1FSIW4a6-USILPpJjPZ-RTM5-dbsboFA91YpLxNR4k1ONZWs/pubhtml}}.

Table \ref{tab:triangleHypothesis} illustrates the state of the hypothesis $h$ for each test execution. The `T' and `F' in each cell represent whether or not the boolean literal in question has been satisfied for that test case (i.e. whether the code element corresponding to the literal has been covered by that test case). Literals that are deleted from $h$ (i.e. evaluate to false for any of the tests executed so far) are shaded. The algorithm starts with $h$ containing the conjunction of all of the boolean literals. This is represented for row $i=0$. Then, for the first input ($i=1$), any boolean literal that does not hold is removed. In this case statements 2 and 3 are executed ($x_{1-3}=T$). Its negation $\overline{x_{1-3}}$ evaluates to false, which means that it is removed from the hypothesis (shaded in the table), etc.

\setlength{\tabcolsep}{0.8pt}
\begin{table}
\caption{Illustration of the inference process, as applied to the triangle example. `T' and `F' indicate whether the element is covered for that given input. Shaded cells indicate that the inference algorithm has deleted that boolean literal from the hypothesis. }
\label{tab:triangleHypothesis}
\tiny
\begin{tabular}{c|ccc||c|c|c|c|c|c|c|c|c|c|c|c|c|c|c|c|c|c|c|c|c|c|c|c|c|c|c|c|c|c|c|c}
$i$&a & b & c & $x_{1-3}$ & $\overline{x_{1-3}}$ & $x_{4}$ & $\overline{x_{4}}$ & $x_{5}$ & $\overline{x_{5}}$ & $x_{6}$ & $\overline{x_{6}}$ & $x_{7}$ & $\overline{x_{7}}$ & $x_{8}$ & $\overline{x_{8}}$ & $x_{9}$ & $\overline{x_{9}}$ & $x_{10}$ & $\overline{x_{10}}$ & $x_{11}$ & $\overline{x_{11}}$ & $x_{14-16}$ & $\overline{x_{14-16}}$ & $x_{17}$ & $\overline{x_{17}}$ & $x_{18-20}$ & $\overline{x_{18-20}}$ & $x_{21}$ & $\overline{x_{21}}$ & $x_{22}$ & $\overline{x_{22}}$ & $x_{23}$ & $\overline{x_{23}}$ & $x_{25}$ & $\overline{x_{25}}$ \\
\hline
0& & & & \cellcolor{white}$\cdot$ & \cellcolor{white}$\cdot$ & \cellcolor{white}$\cdot$ & \cellcolor{white}$\cdot$ & \cellcolor{white}$\cdot$ & \cellcolor{white}$\cdot$ & \cellcolor{white}$\cdot$ & \cellcolor{white}$\cdot$ & \cellcolor{white}$\cdot$ & \cellcolor{white}$\cdot$ & \cellcolor{white}$\cdot$ & \cellcolor{white}$\cdot$ & \cellcolor{white}$\cdot$ & \cellcolor{white}$\cdot$ & \cellcolor{white}$\cdot$ & \cellcolor{white}$\cdot$ & \cellcolor{white}$\cdot$ & \cellcolor{white}$\cdot$ & \cellcolor{white}$\cdot$ & \cellcolor{white}$\cdot$ & \cellcolor{white}$\cdot$ & \cellcolor{white}$\cdot$ & \cellcolor{white}$\cdot$ & \cellcolor{white}$\cdot$ & \cellcolor{white}$\cdot$ & \cellcolor{white}$\cdot$ & \cellcolor{white}$\cdot$ & \cellcolor{white}$\cdot$ & \cellcolor{white}$\cdot$ & \cellcolor{white}$\cdot$ & \cellcolor{white}$\cdot$ & \cellcolor{white}$\cdot$ \\
1 & 1 & 1 & 3 & \cellcolor{white}T & \cellcolor{gray!50}F & \cellcolor{gray!50}F & \cellcolor{white}T & \cellcolor{white}T & \cellcolor{gray!50}F & \cellcolor{gray!50}F & \cellcolor{white}T & \cellcolor{white}T & \cellcolor{gray!50}F & \cellcolor{gray!50}F & \cellcolor{white}T & \cellcolor{white}T & \cellcolor{gray!50}F & \cellcolor{gray!50}F & \cellcolor{white}T & \cellcolor{white}T & \cellcolor{gray!50}F & \cellcolor{gray!50}F & \cellcolor{white}T & \cellcolor{gray!50}F & \cellcolor{white}T & \cellcolor{gray!50}F & \cellcolor{white}T & \cellcolor{gray!50}F & \cellcolor{white}T & \cellcolor{gray!50}F & \cellcolor{white}T & \cellcolor{gray!50}F & \cellcolor{white}T & \cellcolor{white}T & \cellcolor{gray!50}F \\
2 & 5 & 9 & 9 & \cellcolor{white}T & \cellcolor{gray!50}F & \cellcolor{gray!50}F & \cellcolor{white}T & \cellcolor{white}T & \cellcolor{gray!50}F & \cellcolor{gray!50}F & \cellcolor{white}T & \cellcolor{white}T & \cellcolor{gray!50}F & \cellcolor{gray!50}F & \cellcolor{white}T & \cellcolor{white}T & \cellcolor{gray!50}F & \cellcolor{gray!50}T & \cellcolor{gray!50}F & \cellcolor{white}T & \cellcolor{gray!50}F & \cellcolor{gray!50}T & \cellcolor{gray!50}F & \cellcolor{gray!50}F & \cellcolor{white}T & \cellcolor{gray!50}T & \cellcolor{gray!50}F & \cellcolor{gray!50}F & \cellcolor{white}T & \cellcolor{gray!50}T & \cellcolor{gray!50}F & \cellcolor{gray!50}T & \cellcolor{gray!50}F & \cellcolor{white}T & \cellcolor{gray!50}F \\
3 & 4 & 5 & 7 & \cellcolor{white}T & \cellcolor{gray!50}F & \cellcolor{gray!50}T & \cellcolor{gray!50}F & \cellcolor{white}T & \cellcolor{gray!50}F & \cellcolor{gray!50}T & \cellcolor{gray!50}F & \cellcolor{white}T & \cellcolor{gray!50}F & \cellcolor{gray!50}T & \cellcolor{gray!50}F & \cellcolor{white}T & \cellcolor{gray!50}F & \cellcolor{gray!50}T & \cellcolor{gray!50}F & \cellcolor{white}T & \cellcolor{gray!50}F & \cellcolor{gray!50}T & \cellcolor{gray!50}F & \cellcolor{gray!50}F & \cellcolor{white}T & \cellcolor{gray!50}T & \cellcolor{gray!50}F & \cellcolor{gray!50}F & \cellcolor{white}T & \cellcolor{gray!50}T & \cellcolor{gray!50}F & \cellcolor{gray!50}F & \cellcolor{gray!50}T & \cellcolor{white}T & \cellcolor{gray!50}F \\
$\ldots$&\multicolumn{3}{c||}{$\ldots$}&\multicolumn{31}{c}{$\ldots$}\\
199 & 2 & 8 & 8 & \cellcolor{white}T & \cellcolor{gray!50}F & \cellcolor{gray!50}F & \cellcolor{gray!50}T & \cellcolor{white}T & \cellcolor{gray!50}F & \cellcolor{gray!50}T & \cellcolor{gray!50}F & \cellcolor{white}T & \cellcolor{gray!50}F & \cellcolor{gray!50}T & \cellcolor{gray!50}F & \cellcolor{white}T & \cellcolor{gray!50}F & \cellcolor{gray!50}T & \cellcolor{gray!50}F & \cellcolor{white}T & \cellcolor{gray!50}F & \cellcolor{gray!50}T & \cellcolor{gray!50}F & \cellcolor{gray!50}F & \cellcolor{white}T & \cellcolor{gray!50}T & \cellcolor{gray!50}F & \cellcolor{gray!50}F & \cellcolor{gray!50}T & \cellcolor{gray!50}T & \cellcolor{gray!50}F & \cellcolor{gray!50}T & \cellcolor{gray!50}F & \cellcolor{white}T & \cellcolor{gray!50}F \\
\end{tabular}
\end{table}

\paragraph{Hypothesis size}

The most important question for this paper is: how many of these tests do we need to execute in order to guarantee (at least in a PAC context) that the algorithm will finish with an accurate result? And by extension - \emph{how many of these tests do we need to execute to convince ourselves that our test set is inference-adequate}?

For this specific learning problem, we know that $|H_n|=3^n$ \cite{valiant1984theory,kearns1994introduction} (see Sections \ref{subsub:conjunctions} and \ref{sub:inferenceAdequacyAndPAC}). As $n$, we choose the number of theoretically executable lines of code in the system (the actual number is undecidable \cite{weyuker1986axiomatizing}, and this is what the algorithm would seek to infer). To save space, we group statements together that we know sit in the same control-flow block and therefore must be executed together (e.g. $x_{1-3}$)\footnote{This is not a vital part of the process.}. There are $n=16$ blocks, so $|H_{16}|=3^{16}=43,046,721$. For this example we want to guarantee that the result has a maximum error of 10\% ($\epsilon=0.1$), and that the probability of achieving this is 90\% ($\delta = 0.1$). Putting this into Equation \ref{eq:bound}, we get upper bound:

$$m \geq \frac{1}{0.1}(ln(43,046,721) + ln(1/0.1))=198.8038$$

As tests are executed, an increasing number of boolean literals are eliminated from $h$. Most of the eliminations take place within the first few tests. 
By test $i=3$, there is just one boolean literal that is falsely true; the literal $\overline{x_{21}}$ indicates that the line `\texttt{type = EQUILATERAL;}' will never be executed. This is eventually corrected by test $i=112$. By the last test ($i=199$) that we sample (using the upper bound computed above), we end up with the following conjunction: $$x_{1-3} \wedge x_5 \wedge x_7 \wedge x_9 \wedge x_{11} \wedge \overline{x_{17}} \wedge x_{25}$$

This result can be interpreted as follows. Any positive boolean literal (e.g. $x_{1-3}$) corresponds to a statement (or block of statements in this case) that is \emph{always} executed for any execution of the usage profile (i.e. is `relevant' in Miranda and Bertolino's terminology \cite{miranda2015improving}). Any negative literal (e.g. $\overline{x_{17}}$) corresponds to the statement that is not relevant for the usage profile. In this case, $\overline{x_{17}}$ corresponds to the statement that we know can never be executed in our usage context.

Thus we have inferred what Miranda and Bertolino would refer to as the `usage scope' of the triangle program. In our example, the error in our result happens to be 0 (as per our manual inspection). The question then arises, however, to what extent is this a `fluke'? Perhaps our result (which we have inspected and know to be accurate) is merely due to a lucky sample from our input distribution $D_I$. This chance is however factored into our bounds with the $\delta$ parameter (which we set to 0.1). We therefore know that if we repeated this exercise 100 times with different random seeds, in at least 90 of these (the probability represented by $1-\delta$) the test sets that we produce would lead to a model with at least an accuracy of $90\%$ (the accuracy represented by $1-\epsilon$).

It is important to bear in mind that we only illustrate the steps of the inference process to explain the underlying rationale for our adequacy criterion. In practice, no inference or test executions are required and we do not need to actually infer a model. We are instead interested in the theoretical bounds that provide us with the guarantee of accuracy if we were to run this algorithm. 


\subsection{Caveats and Properties}
\label{sub:caveats}

The approach has some interesting properties, but is also subject to a variety of caveats. These are discussed below.

\paragraph{The Relevant Coverage will vary according the input distribution, but the bound on the saturation point will not.} The PAC framework is  `distribution free' \cite{kearns1994introduction}, which means that its results apply to any 
 data distribution. In our context, the specific choice of operational profile used to generate the test inputs does not matter. 
 
 For example, for our triangle example, with each value a, b, and c being sampled from the interval [1,10], the probability of obtaining an input corresponding to an equilateral triangle (where $a=b=c$) is $0.1 * 0.1=0.01$ (this was fortuitously triggered in our example test cases). If, instead, we sampled the inputs from an interval of [1,100], the likelihood of obtaining an equilateral or an isosceles triangle would be much lower (the probability of obtaining an equilateral triangle would be $0.01 * 0.01=0.0001$). Nevertheless, the number of tests for the saturation point remains the same (199), precisely because the likelihood of additional tests covering an additional 10\% of statements (our choice for $\epsilon$) is highly improbable, certainly less than 10\% (our choice for $\delta$).

\paragraph{The approach is agnostic with respect to the logical structure of the source code}

The saturation point for any system can be calculated from only knowing the number of code elements (i.e. statements or branches) in the system, or even just being able to place an upper bound on this number. No source code analysis is required (aside from the ability to count the overall number of statements or branches, depending on the chosen granularity of coverage). The approach does not rely on the actual capability to track test coverage, to encode test executions in a Hit Spectrum, or to infer Boolean Conjunctions. These form the theoretical base that enables us ultimately to justify the use of the formula (Equation \ref{eq:bound}), using the number of coverage targets to compute the hypothesis space.

The logical code structure will of course have an effect on the level of coverage that is achieved at the point of saturation. The execution of a branch or a statement may be subject to combinations of predicates (e.g. controlling the execution of loops or if-statements) that can only be satisfied by highly specific constraints on inputs. Such conditions (such as the code that is executed in the event of an equilateral triangle in our example) may be more or less probable, depending on the input distribution $D_I$. 

Such situations can potentially affect the reliability of our bound. If a condition has a very small probability of being satisfied by inputs in $D_I$, but its execution results in a difference in coverage that is greater than $\epsilon$, this could lead to situations where the bound is unreliable. We suspect that this situation occurs in a small fraction of instances in our evaluation (see Section \ref{sub:results}), and are investigating the implications for the bound as part of our future work.

\paragraph{Assumptions about the input distribution $D_I$}

There are several ways by which to characterize the input distribution $D_I$ in a random testing context. It may be manually constructed in the form of an operational profile \cite{musa1993operational}. It may be implicit in the form of a test generator that can generate quasi-randomized test-inputs, as embodied by random testing frameworks such as QuickCheck \cite{claessen2000quickcheck}.  We use a simple random number generator for the three parameters in our triangle example. For the part of our evaluation that tests object-oriented components, we use a modified version of EvoSuite as a random input generator.

This task of test generation --  of trying to find inputs that will fully exercise the program under test -- is complementary to our task of determining when a point of saturation is reached and we can stop. In principle, given a situation where a test generator produces a larger number of candidate test cases than are feasibly executable, our approach can determine how many tests need to be sampled (at random) to reach saturation. We can, for example, consider a setting where we  know only the input parameters and the number of statements or branches (but do not know how these parameters can affect the program state or the coverage of statements). In this setting, one could opt to use a Combinatorial Testing approach \cite{nie2011survey}. Since these approaches can produce larger numbers of tests than can be feasibly executed, our approach would provide a number that would need to be sampled to reach saturation. 

Regardless of the approach used to generate the distribution of test inputs, it is important that this distribution remains fixed during the test selection process. The act of selecting or generating an input must not change the probability of selecting a subsequent input, because this would  invalidate the `Invariance Assumption' \cite{valiant2014probably} required for PAC guarantees to hold. This constraint means that our approach cannot be used in a feedback-directed way, where tests are generated or selected on the basis of observations made of other tests. 

Finding an approach to computing a PAC bound that \emph{can} involve feedback has been done for certain settings within Machine Learning \cite{angluin1987learning,valiant2014probably}. Recently, the relatively recent emergence of PAC-Bayes \cite{mcallester1998some} (a dynamic generalization of the PAC framework) has enabled the application of bounds to more dynamic learning problems \cite{haddouche2022online}. We are looking into the adaptation of these approaches to accommodate feedback-driven testing techniques, and this forms a part of our future work (Section \ref{sec:conclusions}).

\paragraph{The bound only applies to saturation in terms of code coverage, but does not account for changes in test adequacy that may depend on heap or data-state.}

The accuracy of our approach rests on the assumption that code coverage is a good approximation of tested behavior. In practice, this has to be treated as a significant caveat, and the pitfalls of code coverage have been the subject of a significant amount of research \cite{inozemtseva2014coverage,chekam2017empirical}. Basic code coverage as captured in the Hit Spectrum ignores factors such as data state, or looping behavior.  There are more granular program spectra that subsume the hit spectrum (such as execution count spectra \cite{harrold1998empirical}). However, using these would require a different PAC inference algorithm, and exploring improved spectrum representations and PAC algorithms is part of our future work.                
 
\paragraph{A one-to-many relationship exists between test executions and test objectives}

This is fine if the test objectives in question are traditional code-based targets such as  statements or branches. However, our bound does not apply in situations where test objectives require multiple test cases to be evaluated, as is the case in statistical metamorphic testing for example \cite{guderlei2007statistical}.

One implication of one-to-many relationship between tests and objectives is that we allow for interdependent test objectives. For example, if the execution of one statement is controlled by a predicate, then there is a dependence between the two. This is noteworthy because interdependencies are not permitted by the only other bound on random test sets that we are aware of. Arcuri and Briand's lower-bound \cite{arcuri2011random} requires that test objectives are entirely independent from each other -- i.e. that there must exist a strict one-to-one relationship between tests and objectives.

\section{Evaluation}

In this evaluation we seek to establish the reliability of the bound that is computed, and the relationship to the $\epsilon$ and $\delta$ parameters. Our research questions are as follows:

\begin{itemize}
    \item RQ1: What is the relationship between the parameters $\epsilon$ and $\delta$, the number of test objectives, and the bound on the number of tests?
    \item RQ2: Is the bound reliable with respect to statement coverage?
    \item RQ3: Is the bound reliable with respect to the number of faults that are exposed?
\end{itemize}

\subsection{Methodology}

Our evaluation consists of three studies. The first study focuses on RQ1 and is a functional analysis of the bound function (Equation \ref{eq:bound}) to examine the relationship between the parameters and the bound. The second study explores RQ2 with respect to system tests by examining the relationship between the number of tests executed and the statement coverage on the CARLA driving simulator \cite{dosovitskiy2017carla} and TCP autonomous driving system (ADS) \cite{wu2022TCP}. The third study explores RQs 2 and 3 by establishing the relationship between the number of randomly generated unit tests (as opposed to system tests) and both statement coverage and mutation scores for 7,198 Java classes. We make our experimental materials available online\footnote{ \url{https://anonymous.4open.science/r/RTLT}.}.

\subsubsection{Study 1: Exploring the functional relationship between $\epsilon$ and $\delta$, the number of test objectives, and the bound on the number of tests}

To understand the relationship between the bound and the parameters, we start by considering a fixed number of coverage objectives (used an arbitrary value of 20; although the calculated bound varies for different values, the relationship to $\delta$ and $\epsilon$ does not). We then plot the change in bound for $\delta=[0.01, 0.1, 0.2, 0.3, 0.4, 0.5]$ and $\epsilon=[0.01, 0.1, 0.2, 0.3, 0.4, 0.5]$. We choose values of 0.01 instead of 0 because values must be positive. We then plot the relationship in three dimensions.

To gain an understanding of how the bound scales for different numbers of coverage objectives, we used two configurations: $(\delta=0.1, \epsilon=0.1)$ and a more strict $(\delta=0.05, \epsilon=0.05)$. For each configuration we plot the bound for coverage objectives in the range from 0 to 1,000,000.

\subsubsection{Study 2: Testing an ADS within the CARLA Simulator}

CARLA \cite{dosovitskiy2017carla} is a popular simulation framework implemented in Python to support the development and validation of ADSs. The CARLA GitHub repository\footnote{\url{https://github.com/carla-simulator/carla}} has over 9000 stars and 3000 forks. It presents an interesting case study for us because individual test executions (i.e. simulations of a car driving through a scene) are resource-consuming and can require several minutes to complete. As such, there is a natural motivation to determine the point at which a saturation-point is reached and testing can be terminated.

CARLA provides a range of configurable entities, such as weather conditions, traffic lights, and pedestrians. There is also a leaderboard\footnote{\url{https://leaderboard.carla.org}} to compare the performance of different ADSs, which currently has 30 entries. Submitted ADSs are required to drive from the starting point to the destination point in a set of predefined scenarios. They are then assigned a score based on how much of the route they  completed and any infractions (such as collisions) they committed. 

In this work, we use TCP as our ADS as it was the highest entry on the leaderboard for which code was available and functional at the time we ran our experiments.
For our study, we gathered a set of 2400 executions of TCP by running the default training set for TCP. This consists of 300 routes each for eight of the predefined urban driving environments in CARLA.

For each execution, we gathered statement-wise coverage information for all of the Python files within both CARLA and TCP using the Python \texttt{coverage} module.
We discarded any files where none of the statements were executed by any of the test executions. For each of the remaining 48 files, we calculated the upper-bound on the number of tests using Equation \ref{eq:bound} (we fixed parameters $\delta=0.1$ and $\epsilon=0.2$). We then took a random sub-sample (with replacement) of the test-executions (where the size was determined by the bound) from the 2400 tests, and measured the coverage that was specific to this sub-sample. This was repeated 50 times to accommodate any random variation.

For the results, we calculated the mean coverage, and the upper- and lower confidence intervals. If our upper-bound is reliable (as per RQ2), then coverage should not increase if further tests are added to the test set. To establish this, we compared the coverage for each sub-sample against the coverage that was obtained from all 2400 test executions.

\subsubsection{Study 3: Running random tests on Java units}

We adopted the popular EvoSuite test generation tool \cite{fraser2011evosuite} to act in effect as a random test generator. We do so by disabling all its feedback/guidance capabilities (used for search-based testing). As such it becomes a tool that will generate inputs for Java units that are effectively sampled from a uniform distribution. As our subjects we used a sample of 7,198 Java classes from the SF110 corpus borrowed from a previous large-scale study on unit test generation for Java \cite{fraser2014-SF100}\footnote{\url{https://www.evosuite.org/experimental-data/sf110}}. The SF110 corpus comprises 110 open-source Java projects totaling more than six millions of lines of code. 

We calculated a bound $b$ using Equation \ref{eq:bound}, using the number of statements in the target class as the basis for computing the hypothesis size. We fixed parameters $\delta$ and $\epsilon$ to 0.1. In principle this means that, for a given target, after executing $b$ tests, 90\% of the time, the statement coverage achieved should be within 10\% of what is achievable from that distribution.

To explore whether this is the case (i.e. to answer RQ2), we generated two random test sets for each target with EvoSuite. We generated a test set (which we denote $B$) containing $b$ tests. In order to establish the reliability, we also generated a larger test $L$ set with $2*b$ tests. 

For all of the test sets we gathered statement coverage and mutation coverage metrics. In order to accommodate the randomness in the generation process (and the mutation testing), we repeated each test generation and execution process 10 times on different random seeds.

In order to answer RQ2, we record the proportion of SUTs for which the median difference in statement coverage between $B$ and $L$ is more than 10\%. For the bound to be reliable, this should be the case for a vast majority of cases.

In order to answer RQ3, we repeated the above analyses, but analyzing the mutation coverage instead. Again, we calculate the proportion of systems where the mutation score remains unchanged between $B$ and $L$.

\subsection{Results and Discussion}
\label{sub:results}

\begin{figure}
\setlength{\tabcolsep}{5pt}
\begin{tabular}{p{0.45\textwidth}p{0.45\textwidth}}
   \centering \includegraphics[width=0.3 \textwidth]{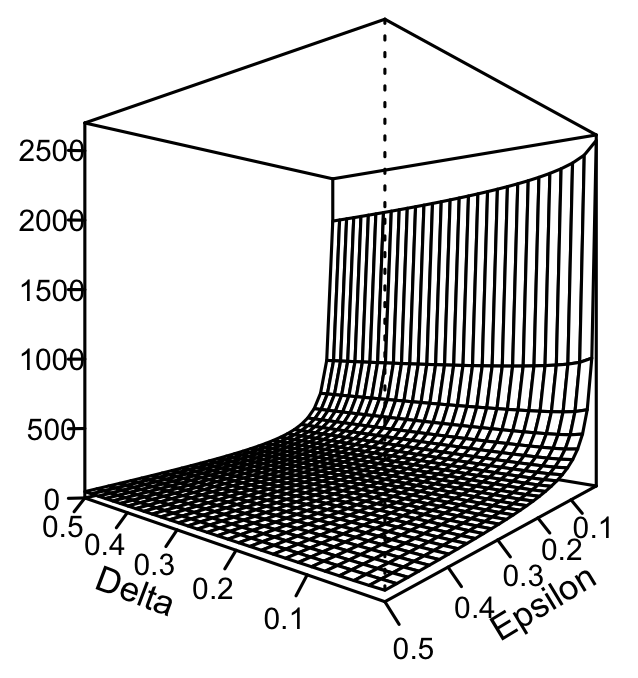}  &  \includegraphics[width = 0.45 \textwidth]{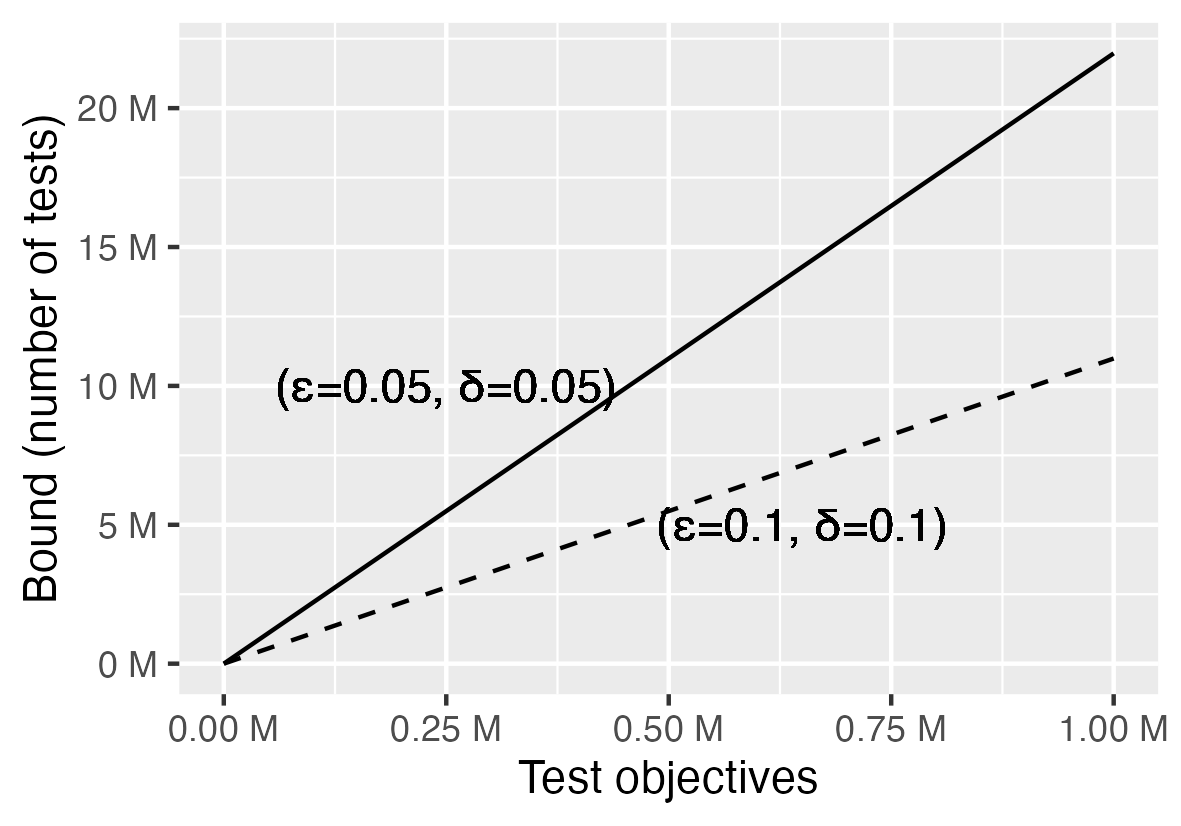}\\
   (a) Relationship between $\delta$, $\epsilon$ and number of tests for a system with 20 test objectives.  & (b) Relationship between number of test objectives and bound on the number of tests.
\end{tabular}

\caption{The relationships between $\delta$, $\epsilon$, the number of test objectives, and the bound on the number of tests.}
\label{fig:rq1}
\end{figure}

\subsubsection*{RQ1: Relationship between $\delta$, $\epsilon$, the number of test objectives, and the bound on the number of tests}

The relationship for a fixed number of test objectives (20) is shown in Figure \ref{fig:rq1}(a). This shows that the number of tests is much more sensitive to $\epsilon$ than it is to $\delta$. There is a sharp increase in the bound as $\epsilon$ tends towards 0. For this reason we only include values in the chart for $\epsilon \geq 0.01$. 

Figure \ref{fig:rq1}(b) shows that, for a given choice of parameters $\delta$ and $\epsilon$, the bound on the number of tests is linear. The gradient of the line reduces for larger values of $\delta$ and (in particular) $\epsilon$. This is to be expected; if we tolerate a larger degree of error in the hypothetical boolean conjunction that would be inferred from the test set, fewer tests would be required by the inference algorithm.

The exponential increase in the number of tests required with respect to $\epsilon$ aligns with the established understanding of the efficacy of random tests. As the test set grows, the number of test cases that are capable of exposing `new' behavior, potentially covering aspects of the program that are as of yet uncovered, and potentially exposing faults, will diminish. The point at which this happens has been referred to as the ``saturation point'' for random testing \cite{lyu1996handbook}. The exponential nature of the decay aligns with findings by B\"ohme and Paul \cite{bohme2015probabilistic}, who showed that the number of faults detected by random testing decays exponentially over time.

It is worth remarking on the size of the bound. The requirement to execute millions of tests (as would be required for non-trivial systems in  Figure \ref{fig:rq1}(b)) may be implausible for many systems -- especially with long execution times.  However, in other cases this is not a problem. For example, at the time of publishing their paper on the random testing of flash storage devices at NASA \cite{groce2007randomized}, the random testing system developed by Groce \emph{et al.} had executed 3.5 billion tests.  

\begin{rqanswer}
\textit{RQ1 Answer: } The bound is strongly affected by $\epsilon$, particularly for $\epsilon < 0.2$. The impact of changing $\delta$ is less than that of $\epsilon$, and is more pronounced for smaller values of $\epsilon$. The bound increases linearly with the number of test objectives for a given configuration of $\epsilon$ and $\delta$.
\end{rqanswer}

\subsubsection*{RQ2: Reliability with respect to statement coverage}

\paragraph{Results from study 2 (CARLA)}

\setlength{\tabcolsep}{6pt}

\begin{table}[t]
\caption{Coverage results for CARLA/TCP. Size refers to number of statements. Bound refers to our upper bound. ``Cov. (all)'' refers to the coverage from running all 2,400 tests. ``Cov. (bound)'' refers to median coverage from running the bounded test sets. $\Delta$ is the difference between ``Cov. (bound)'' and ``Cov. (all)''.}
\label{tab:carla}
\begin{small}
\centering
\begin{tabular}{lcccccc}
  \hline
File & Size & Bound & Cov. (all) & Cov. (bound) & 95\% CI & $\Delta$ \\ 
  \hline
  basic\_agent.py & 128 & 714.62 & 0.1 & 0.1 & [0.1,0.1] & 0 \\ 
  controller.py & 216 & 1198.01 & 0.09 & 0.09 & [0.09,0.09] & 0 \\ 
  global\_route\_planner\_dao.py & 79 & 445.46 & 0.41 & 0.41 & [0.41,0.41] & 0 \\ 
  misc.py & 163 & 906.88 & 0.11 & 0.11 & [0.11,0.11] & 0 \\ 
  autonomous\_agent.py & 126 & 703.64 & 0.29 & 0.29 & [0.29,0.29] & 0 \\ 
  checkpoint\_tools.py & 77 & 434.48 & 0.31 & 0.31 & [0.31,0.31] & 0 \\ 
  result\_writer.py & 116 & 648.71 & 0.53 & 0.53 & [0.53,0.53] & 0 \\ 
  route\_indexer.py & 72 & 407.01 & 0.64 & 0.64 & [0.64,0.64] & 0 \\ 
  route\_manipulation.py & 160 & 890.4 & 0.48 & 0.48 & [0.48,0.48] & 0 \\ 
  planner.py & 113 & 632.23 & 0.57 & 0.57 & [0.57,0.57] & 0 \\ 
  run\_stop\_sign.py & 157 & 873.92 & 0.57 & 0.56 & [0.56,0.56] & 0.01 \\ 
  torch\_layers.py & 114 & 637.72 & 0.31 & 0.31 & [0.31,0.31] & 0 \\ 
  torch\_util.py & 104 & 582.79 & 0.16 & 0.16 & [0.16,0.16] & 0 \\ 
  config\_utils.py & 148 & 824.49 & 0.14 & 0.14 & [0.14,0.14] & 0 \\ 
  rl\_birdview\_wrapper.py & 142 & 791.53 & 0.09 & 0.09 & [0.09,0.09] & 0 \\ 
  traffic\_light.py & 199 & 1104.63 & 0.48 & 0.48 & [0.48,0.48] & 0 \\ 
  transforms.py & 128 & 714.62 & 0.18 & 0.18 & [0.18,0.18] & 0 \\ 
  route\_scenario\_configuration.py & 50 & 286.17 & 0.24 & 0.24 & [0.24,0.24] & 0 \\ 
  scenario\_configuration.py & 86 & 483.92 & 0.5 & 0.5 & [0.5,0.5] & 0 \\ 
  actor\_control.py & 154 & 857.44 & 0.14 & 0.14 & [0.14,0.14] & 0 \\ 
  basic\_control.py & 106 & 593.78 & 0.15 & 0.15 & [0.15,0.15] & 0 \\ 
  external\_control.py & 43 & 247.71 & 0.16 & 0.16 & [0.16,0.16] & 0 \\ 
  npc\_vehicle\_control.py & 106 & 593.78 & 0.13 & 0.13 & [0.13,0.13] & 0 \\ 
  pedestrian\_control.py & 71 & 401.52 & 0.13 & 0.13 & [0.13,0.13] & 0 \\ 
  timer.py & 158 & 879.42 & 0.36 & 0.36 & [0.36,0.36] & 0 \\ 
  traffic\_events.py & 84 & 472.93 & 0.42 & 0.42 & [0.42,0.42] & 0 \\ 
  watchdog.py & 79 & 445.46 & 0.32 & 0.32 & [0.32,0.32] & 0 \\ 
  weather\_sim.py & 166 & 923.36 & 0.19 & 0.19 & [0.19,0.19] & 0 \\ 
  control\_loss.py & 198 & 1099.14 & 0.47 & 0.47 & [0.47,0.47] & 0 \\ 
  junction\_crossing\_route.py & 203 & 1126.6 & 0.14 & 0.14 & [0.14,0.14] & 0 \\ 
  maneuver\_opposite\_direction.py & 172 & 956.32 & 0.1 & 0.1 & [0.1,0.1] & 0 \\ 
  other\_leading\_vehicle.py & 151 & 840.97 & 0.11 & 0.11 & [0.11,0.11] & 0 \\ 
   \hline
\end{tabular}
\end{small}
\end{table}

The results from the CARLA study are summarized in Table \ref{tab:carla}. We restricted our study to files with at least twice as many tests as the bound in our overall sample (we had 2,400 executions of CARLA overall, so we omit any file for which the bound exceeds 1,200). In all cases, the coverage achieved by test sets that are sampled with the size of our bound, and the complete test set (at least twice as large) is the same. The confidence intervals are narrow (the precise coverage value), indicating no variability over the course of the 50 samples per SUT.

\paragraph{Results from study 3 (Java units)}

For the Java units, in 335 cases (4.7\%) of cases, our test generation executions timed-out after 240s, ran out of memory or crashed. We omit these cases from the figures in the rest of our discussion. We compare the coverage between the test sets that were bounded at our upper bound ($b$), and those that contained $2*b$ test cases. The mean difference across all random seeds is 0.3\% (of statements). The 95\% confidence interval for the difference overall is [0.03 0.03]. A Wilcoxon Rank Sum test comparing the numbers of statements covered by the test sets with $b$ and $2*b$ tests returns a $p-$value of 1, indicating no substantive difference in distribution.

\begin{figure}
\begin{tabular}{p{0.45 \textwidth}p{0.45 \textwidth}}
\includegraphics[width=0.45 \textwidth]{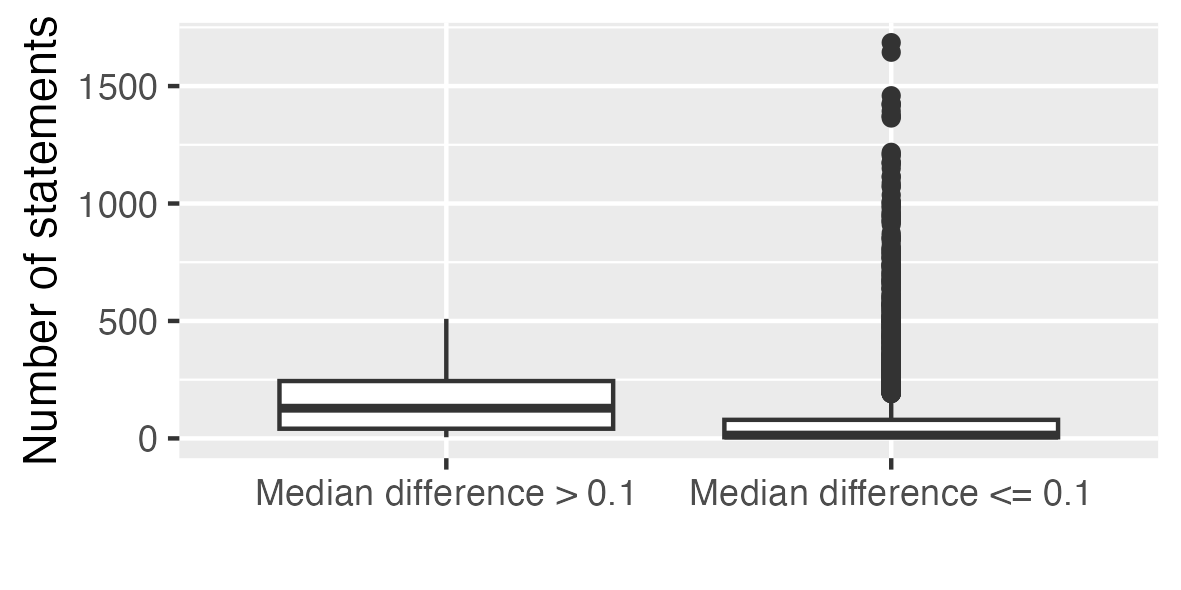}&\includegraphics[width=0.45 \textwidth]{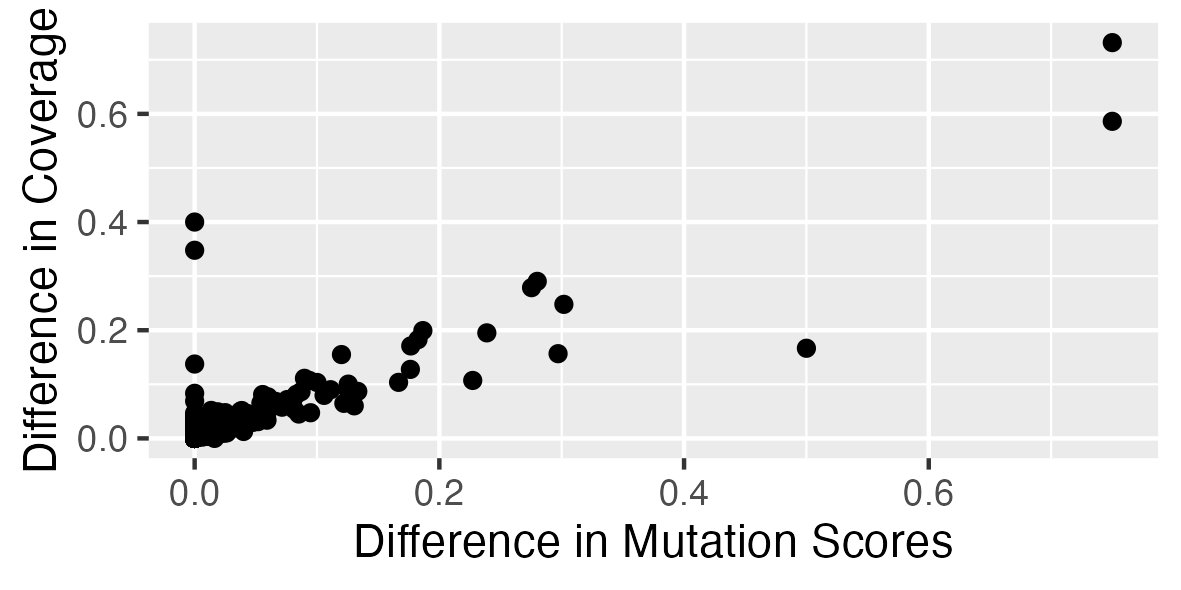}\\
(a) SUT sizes versus coverage difference.&(b) Difference in mutation score versus coverage.\\
\end{tabular}
\caption{Plots of the difference between test sets of size $b$ and $2*b$ with respect to statement coverage and mutation score from Study 3.}
\label{fig:covDifference}
\end{figure}

Since we set $\epsilon$ to 0.1, we would expect the median difference in coverage between the two test sets to be below or equal to 0.1 (i.e. a difference in coverage of less than 10\% of the total number of statements in the SUT)  across all random seeds. This was the case for 99.7\% of the SUTs. 

In 22 SUTs (0.3\% of classes tested), this was not the case. One factor that stands out about these cases (as shown in Figure \ref{fig:covDifference}) is that they tend to be larger than the rest of the classes. For those where the difference is $>0.1$ the median number of coverage targets is 129, whereas for the others the median number is 14. This is however not a determining factor, because the bound holds well for numerous larger classes, up to a maximum of 1685 statements.  

One possible explanation (which we will explore further in future work)  is that these systems may contain significant portions of code that are only executed under specific conditions, fulfilled by a fractionally small proportion of test cases in the input distribution. This would indicate that our bound would be better suited to testing contexts where tests are selected from an operational profile \cite{musa1993operational} than the uniform distribution employed by tools such as EvoSuite.

\begin{rqanswer}
\textit{RQ2 Answer:} The bound is reliable with respect to statement coverage. It held across all of the analyzed CARLA files, as well as 99.7\% of the Java SUTs.
\end{rqanswer}

\subsubsection*{RQ3: Reliability with respect to exposing faults}

In order to examine the reliability with respect to exposing faults, we examine the mutation data gathered during study 3 (Java unit test executions). As was the case with the statement coverage, we compare the mutants exposed for test sets of size $b$ (the computed bound) against test sets of size $2*b$. 

Our results are similar in nature as for RQ2. In 99.6\% of cases, the difference in mutation score between the two test sets was less than 10\%. The 95\% confidence interval for the difference is [0.001,0.002]. This suggests that, for test sets that are larger than the upper bound $b$, the number of faults exposed will tend to remain the same.

In 27 cases (0.4\%) there was however a more significant difference in mutation score. The classes for which this occurs tend to be those classes for which there is also a difference in statement coverage (a relatively strong correlation of 0.87), as shown in Figure \ref{fig:covDifference}(b). This relationship is to be expected, since the execution of source code is a necessary precondition for exposing a mutant. 

\begin{rqanswer}
\textit{RQ3 Answer: }The results indicate that the bound is robust with respect to faults exposed by the test set. Across the SUTs from study 2, the difference in mutation score between bounded test sets and those that were at least twice as large was negligible.
\end{rqanswer}

\subsection{Threats to Validity}

\paragraph{Threats to internal validity}

In both studies 2 and 3, in order to evaluate the reliability of the bound, we compared the coverage achieved against test sets that are at least twice the size of the bound.  An inherent threat is that a larger test set sampled from the same distribution would lead to different levels of coverage (and mutation scores). As part of our future work, we will replicate our studies with larger test sets to evaluate against.

\paragraph{Threats to external validity}

Our studies focused on two specific types of system (Java units and an automated driving system), with respect to two specific types of test distribution (selected through EvoSuite and the TCP test suite respectively). Although our results are consistent across these two very different contexts, in our future work we will investigate other types of systems to ensure that the results generalize.

Our conclusions from RQ3 (reliability with respect to the ability to expose faults) are based on the use of mutation testing. For this we used the mutation engine built into EvoSuite. This gives rise to the threat that the mutants used here are not reflective of genuine faults. Our future work will focus on replicating this study on curated sets of faults, such as Defects4J \cite{just2014defects4j}. 

\section{Related Work}
\label{sec:relatedWork}




 Goldreich \emph{et al.} \cite{goldreich1998property} investigated the relationship between PAC-learning and `property testing'. They consider a specific graph-theoretical context, where the program in question manipulates a graph and the goal is to check that the output conforms to particular properties (e.g. is bipartite, etc.). They adopt a PAC-learning inspired approach to place limits on the number of tests that would be required to offer statistical guarantees that the graphs adhere to particular properties.

 Chen \emph{et al.} \cite{chen2016pac} modified  Angluin's $L*$ algorithm to infer automata of software systems for formal verification \cite{angluin1987learning}. They exploit the fact that $L*$ is a PAC algorithm to make statistically justifiable guarantees about the correctness of the properties that they verify, and are able to apply Angluin's original PAC upper-bound to the number of tests required. They present a good example of how PAC bounds can be used for test bounds, but are restricted to the specific situation where the system and its correctness properties can be modeled as a finite state automata. 
 
  As discussed in Section \ref{sub:inferenceAdequacy}, there have been several attempts to relate inference adequacy to PAC. In addition to the more theoretical contributions made by Zhu \emph{et al.} \cite{zhu1992inductive,zhu1996formal} and Romanik and Vitter \cite{romanik1993using,romanik1996using} (discussed previously), there has also been work by Fraser and Walkinshaw to apply PAC inference adequacy in an entirely empirical context \cite{fraser2015assessing}. They do not use PAC bounds to establish a limit on the number of tests required.
  
  One inherent limitation of the approaches discussed above is the need to characterize testing complexity in terms of input and output. Although this can work well for programs with input-output behaviors that are well understood, such as real numbers or graphs, this can become challenging to reason with when the input or output space involves complex types, or where the types of rules that govern the relationship between input- and output are not known, or are not known to be PAC-learnable. Consequently, the underlying complexity measure for a program can be difficult to accurately compute \cite{romanik1996using}, which means that the test set is difficult to bound. 

 In contrast, our approach conceptualizes the test execution behavior of a program in terms of the code coverage spectrum. This means that we are creating a direct link to recognized test adequacy concepts \cite{bertolino2019adaptive,miranda2020testing}. It also means that we can characterize the PAC-learning problem as the task of learning Boolean conjunctions, which provides us with our ability to produce bounds on test set size that are computed from a finite hypothesis space, and these bounds tend to be much tighter \cite{kearns1994introduction} than those that are computed using the Vapnik-Chervonenkis dimension \cite{vapnik1999nature}. 

\section{Conclusions and Future Work}
\label{sec:conclusions}

We have shown that the saturation point -- the number of tests beyond which there is a negligible improvement in test coverage -- can be bounded. This can be achieved from only knowing the number of test goals (i.e. statements or branches) in the system under test. This provides an answer to one of the fundamental questions in software testing of when to stop \cite{frankl1988applicable}. Our studies indicate that this bound is reliable, both in terms of statement coverage and the ability to expose faults. 


In the context of fuzzing, B\"ohme \cite{bohme2019assurances} points out that there is a lack of techniques that enable testers to decide whether or not to stop testing. Questions around residual risk (the likelihood of faults being uncovered by further tests) and the cost-benefit (the potential reward in running more tests) are crucial. Although we do not deal with the situation of fuzzing in this paper (see below), these questions nevertheless apply in any random testing context. We have shown how the tester can obtain an answer to these questions. Crucially, this answer can be established \emph{before} any tests have been executed, just from knowing the number of statements or branches in the source code.

The use of this bound is subject to several caveats (see Section \ref{sub:caveats}). Firstly, the distribution of tests must be fixed, because PAC relies on the distribution ``Invariance Assumption'' \cite{valiant2014probably}. Though fine for conventional random testing, it rules out feedback-directed testing approaches such as evolutionary testing and fuzzing. Recently, progress has been made to generalize the PAC framework to allow for certain transformations in the distribution \cite{shao2022theory}. We will work to adapt these advances into a testing context, to enable us to reason about test adequacy in a feedback-directed testing context.

A further noteworthy caveat of our technique is that it is tied to representing program behavior in terms of a Hit Spectrum (because of the direct link to the PAC-learnable Boolean conjunctions). Recently  however, the emergence of the PAC-Bayes framework \cite{mcallester1998some} has enabled the application of PAC to algorithms that output distributions as opposed to simple classifications. This has enabled bounds to be produced for a much broader range of learners including deep neural nets \cite{DBLP:journals/corr/DziugaiteR17}, and causal graph discovery \cite{kyono2020castle}. Our future work will investigate the application of these more expressive types of models to offer a more general bound on random test set size.  

\begin{acks}
Walkinshaw, Foster and Hierons are sponsored by the  \grantsponsor{EPSRC}{EPSRC}{https://www.ukri.org/councils/epsrc/} under the \grantnum{}{CITCOM project \\EP/T030526/1}. We thank the reviewers and Seongmin Lee for their helpful feedback.
\end{acks}

\bibliographystyle{ACM-Reference-Format}
\bibliography{sample-base}


\begin{thebibliography}{68}


\ifx \showCODEN    \undefined \def \showCODEN     #1{\unskip}     \fi
\ifx \showDOI      \undefined \def \showDOI       #1{#1}\fi
\ifx \showISBNx    \undefined \def \showISBNx     #1{\unskip}     \fi
\ifx \showISBNxiii \undefined \def \showISBNxiii  #1{\unskip}     \fi
\ifx \showISSN     \undefined \def \showISSN      #1{\unskip}     \fi
\ifx \showLCCN     \undefined \def \showLCCN      #1{\unskip}     \fi
\ifx \shownote     \undefined \def \shownote      #1{#1}          \fi
\ifx \showarticletitle \undefined \def \showarticletitle #1{#1}   \fi
\ifx \showURL      \undefined \def \showURL       {\relax}        \fi
\providecommand\bibfield[2]{#2}
\providecommand\bibinfo[2]{#2}
\providecommand\natexlab[1]{#1}
\providecommand\showeprint[2][]{arXiv:#2}

\bibitem[Amalfitano et~al\mbox{.}(2015)]%
        {amalfitano2015exploiting}
\bibfield{author}{\bibinfo{person}{Domenico Amalfitano},
  \bibinfo{person}{Nicola Amatucci}, \bibinfo{person}{Anna~Rita Fasolino},
  \bibinfo{person}{Porfirio Tramontana}, \bibinfo{person}{Emily Kowalczyk},
  {and} \bibinfo{person}{Atif~M Memon}.} \bibinfo{year}{2015}\natexlab{}.
\newblock \showarticletitle{Exploiting the saturation effect in automatic
  random testing of android applications}. In \bibinfo{booktitle}{\emph{2015
  2nd ACM International Conference on Mobile Software Engineering and
  Systems}}. IEEE, \bibinfo{pages}{33--43}.
\newblock


\bibitem[Angluin(1987)]%
        {angluin1987learning}
\bibfield{author}{\bibinfo{person}{Dana Angluin}.}
  \bibinfo{year}{1987}\natexlab{}.
\newblock \showarticletitle{Learning regular sets from queries and
  counterexamples}.
\newblock \bibinfo{journal}{\emph{Information and computation}}
  \bibinfo{volume}{75}, \bibinfo{number}{2} (\bibinfo{year}{1987}),
  \bibinfo{pages}{87--106}.
\newblock


\bibitem[Arcuri et~al\mbox{.}(2011)]%
        {arcuri2011random}
\bibfield{author}{\bibinfo{person}{Andrea Arcuri},
  \bibinfo{person}{Muhammad~Zohaib Iqbal}, {and} \bibinfo{person}{Lionel
  Briand}.} \bibinfo{year}{2011}\natexlab{}.
\newblock \showarticletitle{Random testing: Theoretical results and practical
  implications}.
\newblock \bibinfo{journal}{\emph{IEEE transactions on Software Engineering}}
  \bibinfo{volume}{38}, \bibinfo{number}{2} (\bibinfo{year}{2011}),
  \bibinfo{pages}{258--277}.
\newblock


\bibitem[Bergadano and Gunetti(1996)]%
        {bergadano1996testing}
\bibfield{author}{\bibinfo{person}{Francesco Bergadano} {and}
  \bibinfo{person}{Daniele Gunetti}.} \bibinfo{year}{1996}\natexlab{}.
\newblock \showarticletitle{Testing by means of inductive program learning}.
\newblock \bibinfo{journal}{\emph{ACM Transactions on Software Engineering and
  Methodology (TOSEM)}} \bibinfo{volume}{5}, \bibinfo{number}{2}
  (\bibinfo{year}{1996}), \bibinfo{pages}{119--145}.
\newblock


\bibitem[Bertolino et~al\mbox{.}(2019)]%
        {bertolino2019adaptive}
\bibfield{author}{\bibinfo{person}{Antonia Bertolino}, \bibinfo{person}{Breno
  Miranda}, \bibinfo{person}{Roberto Pietrantuono}, {and}
  \bibinfo{person}{Stefano Russo}.} \bibinfo{year}{2019}\natexlab{}.
\newblock \showarticletitle{Adaptive test case allocation, selection and
  generation using coverage spectrum and operational profile}.
\newblock \bibinfo{journal}{\emph{IEEE Transactions on Software Engineering}}
  \bibinfo{volume}{47}, \bibinfo{number}{5} (\bibinfo{year}{2019}),
  \bibinfo{pages}{881--898}.
\newblock


\bibitem[B{\"o}hme(2019)]%
        {bohme2019assurances}
\bibfield{author}{\bibinfo{person}{Marcel B{\"o}hme}.}
  \bibinfo{year}{2019}\natexlab{}.
\newblock \showarticletitle{Assurances in software testing: A roadmap}. In
  \bibinfo{booktitle}{\emph{2019 IEEE/ACM 41st International Conference on
  Software Engineering: New Ideas and Emerging Results (ICSE-NIER)}}. IEEE,
  \bibinfo{pages}{5--8}.
\newblock


\bibitem[B{\"o}hme and Paul(2015)]%
        {bohme2015probabilistic}
\bibfield{author}{\bibinfo{person}{Marcel B{\"o}hme} {and}
  \bibinfo{person}{Soumya Paul}.} \bibinfo{year}{2015}\natexlab{}.
\newblock \showarticletitle{A probabilistic analysis of the efficiency of
  automated software testing}.
\newblock \bibinfo{journal}{\emph{IEEE Transactions on Software Engineering}}
  \bibinfo{volume}{42}, \bibinfo{number}{4} (\bibinfo{year}{2015}),
  \bibinfo{pages}{345--360}.
\newblock


\bibitem[Briand et~al\mbox{.}(2009)]%
        {briand2009using}
\bibfield{author}{\bibinfo{person}{Lionel~C Briand}, \bibinfo{person}{Yvan
  Labiche}, \bibinfo{person}{Zaheer Bawar}, {and} \bibinfo{person}{Nadia~Traldi
  Spido}.} \bibinfo{year}{2009}\natexlab{}.
\newblock \showarticletitle{Using machine learning to refine category-partition
  test specifications and test suites}.
\newblock \bibinfo{journal}{\emph{Information and Software Technology}}
  \bibinfo{volume}{51}, \bibinfo{number}{11} (\bibinfo{year}{2009}),
  \bibinfo{pages}{1551--1564}.
\newblock


\bibitem[Budd and Angluin(1982)]%
        {budd1982two}
\bibfield{author}{\bibinfo{person}{Timothy~A Budd} {and} \bibinfo{person}{Dana
  Angluin}.} \bibinfo{year}{1982}\natexlab{}.
\newblock \showarticletitle{Two notions of correctness and their relation to
  testing}.
\newblock \bibinfo{journal}{\emph{Acta informatica}}  \bibinfo{volume}{18}
  (\bibinfo{year}{1982}), \bibinfo{pages}{31--45}.
\newblock


\bibitem[Chekam et~al\mbox{.}(2017)]%
        {chekam2017empirical}
\bibfield{author}{\bibinfo{person}{Thierry~Titcheu Chekam},
  \bibinfo{person}{Mike Papadakis}, \bibinfo{person}{Yves Le~Traon}, {and}
  \bibinfo{person}{Mark Harman}.} \bibinfo{year}{2017}\natexlab{}.
\newblock \showarticletitle{An empirical study on mutation, statement and
  branch coverage fault revelation that avoids the unreliable clean program
  assumption}. In \bibinfo{booktitle}{\emph{2017 IEEE/ACM 39th International
  Conference on Software Engineering (ICSE)}}. IEEE, \bibinfo{pages}{597--608}.
\newblock


\bibitem[Chen et~al\mbox{.}(2005)]%
        {chen2005adaptive}
\bibfield{author}{\bibinfo{person}{Tsong~Yueh Chen}, \bibinfo{person}{Hing
  Leung}, {and} \bibinfo{person}{Ieng~Kei Mak}.}
  \bibinfo{year}{2005}\natexlab{}.
\newblock \showarticletitle{Adaptive random testing}. In
  \bibinfo{booktitle}{\emph{Advances in Computer Science-ASIAN 2004.
  Higher-Level Decision Making: 9th Asian Computing Science Conference.}}
  Springer, \bibinfo{pages}{320--329}.
\newblock


\bibitem[Chen et~al\mbox{.}(2016)]%
        {chen2016pac}
\bibfield{author}{\bibinfo{person}{Yu-Fang Chen}, \bibinfo{person}{Chiao
  Hsieh}, \bibinfo{person}{Ond{\v{r}}ej Leng{\'a}l}, \bibinfo{person}{Tsung-Ju
  Lii}, \bibinfo{person}{Ming-Hsien Tsai}, \bibinfo{person}{Bow-Yaw Wang},
  {and} \bibinfo{person}{Farn Wang}.} \bibinfo{year}{2016}\natexlab{}.
\newblock \showarticletitle{PAC learning-based verification and model
  synthesis}. In \bibinfo{booktitle}{\emph{Proceedings of the 38th
  International Conference on Software Engineering}}.
  \bibinfo{pages}{714--724}.
\newblock


\bibitem[Choi et~al\mbox{.}(2013)]%
        {choi2013guided}
\bibfield{author}{\bibinfo{person}{Wontae Choi}, \bibinfo{person}{George
  Necula}, {and} \bibinfo{person}{Koushik Sen}.}
  \bibinfo{year}{2013}\natexlab{}.
\newblock \showarticletitle{Guided gui testing of android apps with minimal
  restart and approximate learning}.
\newblock \bibinfo{journal}{\emph{Acm Sigplan Notices}} \bibinfo{volume}{48},
  \bibinfo{number}{10} (\bibinfo{year}{2013}), \bibinfo{pages}{623--640}.
\newblock


\bibitem[Claessen and Hughes(2000)]%
        {claessen2000quickcheck}
\bibfield{author}{\bibinfo{person}{Koen Claessen} {and} \bibinfo{person}{John
  Hughes}.} \bibinfo{year}{2000}\natexlab{}.
\newblock \showarticletitle{QuickCheck: a lightweight tool for random testing
  of Haskell programs}. In \bibinfo{booktitle}{\emph{Proceedings of the fifth
  ACM SIGPLAN international conference on Functional programming}}.
  \bibinfo{pages}{268--279}.
\newblock


\bibitem[Dosovitskiy et~al\mbox{.}(2017)]%
        {dosovitskiy2017carla}
\bibfield{author}{\bibinfo{person}{Alexey Dosovitskiy}, \bibinfo{person}{German
  Ros}, \bibinfo{person}{Felipe Codevilla}, \bibinfo{person}{Antonio Lopez},
  {and} \bibinfo{person}{Vladlen Koltun}.} \bibinfo{year}{2017}\natexlab{}.
\newblock \showarticletitle{CARLA: An open urban driving simulator}. In
  \bibinfo{booktitle}{\emph{Conference on robot learning}}. PMLR,
  \bibinfo{pages}{1--16}.
\newblock


\bibitem[Dupont et~al\mbox{.}(2008)]%
        {dupont2008qsm}
\bibfield{author}{\bibinfo{person}{Pierre Dupont}, \bibinfo{person}{Bernard
  Lambeau}, \bibinfo{person}{Christophe Damas}, {and} \bibinfo{person}{Axel~van
  Lamsweerde}.} \bibinfo{year}{2008}\natexlab{}.
\newblock \showarticletitle{The QSM algorithm and its application to software
  behavior model induction}.
\newblock \bibinfo{journal}{\emph{Applied Artificial Intelligence}}
  \bibinfo{volume}{22}, \bibinfo{number}{1-2} (\bibinfo{year}{2008}),
  \bibinfo{pages}{77--115}.
\newblock


\bibitem[Dziugaite and Roy(2017)]%
        {DBLP:journals/corr/DziugaiteR17}
\bibfield{author}{\bibinfo{person}{Gintare~Karolina Dziugaite} {and}
  \bibinfo{person}{Daniel~M. Roy}.} \bibinfo{year}{2017}\natexlab{}.
\newblock \showarticletitle{Computing Nonvacuous Generalization Bounds for Deep
  (Stochastic) Neural Networks with Many More Parameters than Training Data}.
\newblock \bibinfo{journal}{\emph{CoRR}}  \bibinfo{volume}{abs/1703.11008}
  (\bibinfo{year}{2017}).
\newblock
\showeprint[arXiv]{1703.11008}
\urldef\tempurl%
\url{https://arxiv.org/abs/1703.11008}
\showURL{%
\tempurl}


\bibitem[Frankl and Weyuker(1988)]%
        {frankl1988applicable}
\bibfield{author}{\bibinfo{person}{Phyllis~G. Frankl} {and}
  \bibinfo{person}{Elaine~J. Weyuker}.} \bibinfo{year}{1988}\natexlab{}.
\newblock \showarticletitle{An applicable family of data flow testing
  criteria}.
\newblock \bibinfo{journal}{\emph{IEEE Transactions on Software Engineering}}
  \bibinfo{volume}{14}, \bibinfo{number}{10} (\bibinfo{year}{1988}),
  \bibinfo{pages}{1483--1498}.
\newblock


\bibitem[Fraser and Arcuri(2011)]%
        {fraser2011evosuite}
\bibfield{author}{\bibinfo{person}{Gordon Fraser} {and} \bibinfo{person}{Andrea
  Arcuri}.} \bibinfo{year}{2011}\natexlab{}.
\newblock \showarticletitle{Evosuite: automatic test suite generation for
  object-oriented software}. In \bibinfo{booktitle}{\emph{Proceedings of the
  19th ACM SIGSOFT symposium and the 13th European conference on Foundations of
  software engineering}}. \bibinfo{pages}{416--419}.
\newblock


\bibitem[Fraser and Arcuri(2014)]%
        {fraser2014-SF100}
\bibfield{author}{\bibinfo{person}{Gordon Fraser} {and} \bibinfo{person}{Andrea
  Arcuri}.} \bibinfo{year}{2014}\natexlab{}.
\newblock \showarticletitle{A Large Scale Evaluation of Automated Unit Test
  Generation Using EvoSuite}.
\newblock \bibinfo{journal}{\emph{ACM Transactions on Software Engineering and
  Methodology (TOSEM)}} \bibinfo{volume}{24}, \bibinfo{number}{2}
  (\bibinfo{year}{2014}), \bibinfo{pages}{8}.
\newblock


\bibitem[Fraser and Walkinshaw(2015)]%
        {fraser2015assessing}
\bibfield{author}{\bibinfo{person}{Gordon Fraser} {and} \bibinfo{person}{Neil
  Walkinshaw}.} \bibinfo{year}{2015}\natexlab{}.
\newblock \showarticletitle{Assessing and generating test sets in terms of
  behavioural adequacy}.
\newblock \bibinfo{journal}{\emph{Software Testing, Verification and
  Reliability}} \bibinfo{volume}{25}, \bibinfo{number}{8}
  (\bibinfo{year}{2015}), \bibinfo{pages}{749--780}.
\newblock


\bibitem[Ghani and Clark(2008)]%
        {ghani2008strengthening}
\bibfield{author}{\bibinfo{person}{Kamran Ghani} {and} \bibinfo{person}{John~A
  Clark}.} \bibinfo{year}{2008}\natexlab{}.
\newblock \showarticletitle{Strengthening inferred specifications using search
  based testing}. In \bibinfo{booktitle}{\emph{2008 IEEE International
  Conference on Software Testing Verification and Validation Workshop}}. IEEE,
  \bibinfo{pages}{187--194}.
\newblock


\bibitem[Gold(1978)]%
        {gold1978complexity}
\bibfield{author}{\bibinfo{person}{E~Mark Gold}.}
  \bibinfo{year}{1978}\natexlab{}.
\newblock \showarticletitle{Complexity of automaton identification from given
  data}.
\newblock \bibinfo{journal}{\emph{Information and control}}
  \bibinfo{volume}{37}, \bibinfo{number}{3} (\bibinfo{year}{1978}),
  \bibinfo{pages}{302--320}.
\newblock


\bibitem[Goldreich et~al\mbox{.}(1998)]%
        {goldreich1998property}
\bibfield{author}{\bibinfo{person}{Oded Goldreich}, \bibinfo{person}{Shari
  Goldwasser}, {and} \bibinfo{person}{Dana Ron}.}
  \bibinfo{year}{1998}\natexlab{}.
\newblock \showarticletitle{Property testing and its connection to learning and
  approximation}.
\newblock \bibinfo{journal}{\emph{Journal of the ACM (JACM)}}
  \bibinfo{volume}{45}, \bibinfo{number}{4} (\bibinfo{year}{1998}),
  \bibinfo{pages}{653--750}.
\newblock


\bibitem[Goodenough and Gerhart(1975)]%
        {goodenough1975toward}
\bibfield{author}{\bibinfo{person}{John~B Goodenough} {and}
  \bibinfo{person}{Susan~L Gerhart}.} \bibinfo{year}{1975}\natexlab{}.
\newblock \showarticletitle{Toward a theory of test data selection}. In
  \bibinfo{booktitle}{\emph{Proceedings of the international conference on
  Reliable software}}. \bibinfo{pages}{493--510}.
\newblock


\bibitem[Groce et~al\mbox{.}(2007)]%
        {groce2007randomized}
\bibfield{author}{\bibinfo{person}{Alex Groce}, \bibinfo{person}{Gerard
  Holzmann}, {and} \bibinfo{person}{Rajeev Joshi}.}
  \bibinfo{year}{2007}\natexlab{}.
\newblock \showarticletitle{Randomized differential testing as a prelude to
  formal verification}. In \bibinfo{booktitle}{\emph{29th International
  Conference on Software Engineering (ICSE'07)}}. IEEE,
  \bibinfo{pages}{621--631}.
\newblock


\bibitem[Guderlei and Mayer(2007)]%
        {guderlei2007statistical}
\bibfield{author}{\bibinfo{person}{Ralph Guderlei} {and}
  \bibinfo{person}{Johannes Mayer}.} \bibinfo{year}{2007}\natexlab{}.
\newblock \showarticletitle{Statistical metamorphic testing testing programs
  with random output by means of statistical hypothesis tests and metamorphic
  testing}. In \bibinfo{booktitle}{\emph{Seventh International Conference on
  Quality Software (QSIC 2007)}}. IEEE, \bibinfo{pages}{404--409}.
\newblock


\bibitem[Haddouche and Guedj(2022)]%
        {haddouche2022online}
\bibfield{author}{\bibinfo{person}{Maxime Haddouche} {and}
  \bibinfo{person}{Benjamin Guedj}.} \bibinfo{year}{2022}\natexlab{}.
\newblock \showarticletitle{Online pac-bayes learning}.
\newblock \bibinfo{journal}{\emph{Advances in Neural Information Processing
  Systems}}  \bibinfo{volume}{35} (\bibinfo{year}{2022}),
  \bibinfo{pages}{25725--25738}.
\newblock


\bibitem[Hamlet(1994)]%
        {hamlet1994}
\bibfield{author}{\bibinfo{person}{Richard Hamlet}.}
  \bibinfo{year}{1994}\natexlab{}.
\newblock \showarticletitle{Random Testing}.
\newblock \bibinfo{journal}{\emph{Encyclopedia of Software Engineering}}
  (\bibinfo{year}{1994}).
\newblock


\bibitem[Harrold(2000)]%
        {harrold2000testing}
\bibfield{author}{\bibinfo{person}{Mary~Jean Harrold}.}
  \bibinfo{year}{2000}\natexlab{}.
\newblock \showarticletitle{Testing: a roadmap}. In
  \bibinfo{booktitle}{\emph{Proceedings of the Conference on the Future of
  Software Engineering}}. \bibinfo{pages}{61--72}.
\newblock


\bibitem[Harrold et~al\mbox{.}(1998)]%
        {harrold1998empirical}
\bibfield{author}{\bibinfo{person}{Mary~Jean Harrold}, \bibinfo{person}{Gregg
  Rothermel}, \bibinfo{person}{Rui Wu}, {and} \bibinfo{person}{Liu Yi}.}
  \bibinfo{year}{1998}\natexlab{}.
\newblock \showarticletitle{An empirical investigation of program spectra}. In
  \bibinfo{booktitle}{\emph{Proceedings of the 1998 ACM SIGPLAN-SIGSOFT
  workshop on Program analysis for software tools and engineering}}.
  \bibinfo{pages}{83--90}.
\newblock


\bibitem[Haussler(1988)]%
        {haussler1988quantifying}
\bibfield{author}{\bibinfo{person}{David Haussler}.}
  \bibinfo{year}{1988}\natexlab{}.
\newblock \showarticletitle{Quantifying inductive bias: AI learning algorithms
  and Valiant's learning framework}.
\newblock \bibinfo{journal}{\emph{Artificial intelligence}}
  \bibinfo{volume}{36}, \bibinfo{number}{2} (\bibinfo{year}{1988}),
  \bibinfo{pages}{177--221}.
\newblock


\bibitem[Inozemtseva and Holmes(2014)]%
        {inozemtseva2014coverage}
\bibfield{author}{\bibinfo{person}{Laura Inozemtseva} {and}
  \bibinfo{person}{Reid Holmes}.} \bibinfo{year}{2014}\natexlab{}.
\newblock \showarticletitle{Coverage is not strongly correlated with test suite
  effectiveness}. In \bibinfo{booktitle}{\emph{Proceedings of the 36th
  international conference on software engineering}}.
  \bibinfo{pages}{435--445}.
\newblock


\bibitem[Isberner et~al\mbox{.}(2015)]%
        {isberner2015open}
\bibfield{author}{\bibinfo{person}{Malte Isberner}, \bibinfo{person}{Falk
  Howar}, {and} \bibinfo{person}{Bernhard Steffen}.}
  \bibinfo{year}{2015}\natexlab{}.
\newblock \showarticletitle{The open-source learnLib: a framework for active
  automata learning}. In \bibinfo{booktitle}{\emph{Computer Aided Verification:
  27th International Conference, CAV 2015, San Francisco, CA, USA, July 18-24,
  2015, Proceedings, Part I 27}}. Springer, \bibinfo{pages}{487--495}.
\newblock


\bibitem[Ivankovi{\'c} et~al\mbox{.}(2019)]%
        {ivankovic2019code}
\bibfield{author}{\bibinfo{person}{Marko Ivankovi{\'c}}, \bibinfo{person}{Goran
  Petrovi{\'c}}, \bibinfo{person}{Ren{\'e} Just}, {and} \bibinfo{person}{Gordon
  Fraser}.} \bibinfo{year}{2019}\natexlab{}.
\newblock \showarticletitle{Code coverage at Google}. In
  \bibinfo{booktitle}{\emph{Proceedings of the 2019 27th ACM Joint Meeting on
  European Software Engineering Conference and Symposium on the Foundations of
  Software Engineering}}. \bibinfo{pages}{955--963}.
\newblock


\bibitem[Just et~al\mbox{.}(2014)]%
        {just2014defects4j}
\bibfield{author}{\bibinfo{person}{Ren{\'e} Just}, \bibinfo{person}{Darioush
  Jalali}, {and} \bibinfo{person}{Michael~D Ernst}.}
  \bibinfo{year}{2014}\natexlab{}.
\newblock \showarticletitle{Defects4J: A database of existing faults to enable
  controlled testing studies for Java programs}. In
  \bibinfo{booktitle}{\emph{Proceedings of the 2014 international symposium on
  software testing and analysis}}. \bibinfo{pages}{437--440}.
\newblock


\bibitem[Kearns and Vazirani(1994)]%
        {kearns1994introduction}
\bibfield{author}{\bibinfo{person}{Michael~J Kearns} {and}
  \bibinfo{person}{Umesh Vazirani}.} \bibinfo{year}{1994}\natexlab{}.
\newblock \bibinfo{booktitle}{\emph{An introduction to computational learning
  theory}}.
\newblock \bibinfo{publisher}{MIT press}.
\newblock


\bibitem[Kyono et~al\mbox{.}(2020)]%
        {kyono2020castle}
\bibfield{author}{\bibinfo{person}{Trent Kyono}, \bibinfo{person}{Yao Zhang},
  {and} \bibinfo{person}{Mihaela van~der Schaar}.}
  \bibinfo{year}{2020}\natexlab{}.
\newblock \showarticletitle{Castle: Regularization via auxiliary causal graph
  discovery}.
\newblock \bibinfo{journal}{\emph{Advances in Neural Information Processing
  Systems}}  \bibinfo{volume}{33} (\bibinfo{year}{2020}),
  \bibinfo{pages}{1501--1512}.
\newblock


\bibitem[Liyanage et~al\mbox{.}(2023)]%
        {liyanage2023reachable}
\bibfield{author}{\bibinfo{person}{Danushka Liyanage}, \bibinfo{person}{Marcel
  B{\"o}hme}, \bibinfo{person}{Chakkrit Tantithamthavorn}, {and}
  \bibinfo{person}{Stephan Lipp}.} \bibinfo{year}{2023}\natexlab{}.
\newblock \showarticletitle{Reachable Coverage: Estimating Saturation in
  Fuzzing}. In \bibinfo{booktitle}{\emph{Proceedings of the 45th IEEE/ACM
  International Conference on Software Engineering (ICSE'23), 17-19 May 2023,
  Australia}}.
\newblock


\bibitem[Liyanage et~al\mbox{.}(2024)]%
        {liyanage2024extrapolating}
\bibfield{author}{\bibinfo{person}{Danushka Liyanage},
  \bibinfo{person}{Seongmin Lee}, \bibinfo{person}{Chakkrit Tantithamthavorn},
  {and} \bibinfo{person}{Marcel B{\"o}hme}.} \bibinfo{year}{2024}\natexlab{}.
\newblock \showarticletitle{Extrapolating Coverage Rate in Greybox Fuzzing}. In
  \bibinfo{booktitle}{\emph{Proceedings of the IEEE/ACM 46th International
  Conference on Software Engineering}}. \bibinfo{pages}{1--12}.
\newblock


\bibitem[Lyu et~al\mbox{.}(1996)]%
        {lyu1996handbook}
\bibfield{author}{\bibinfo{person}{Michael~R Lyu} {et~al\mbox{.}}}
  \bibinfo{year}{1996}\natexlab{}.
\newblock \bibinfo{booktitle}{\emph{Handbook of software reliability
  engineering}}. Vol.~\bibinfo{volume}{222}.
\newblock \bibinfo{publisher}{IEEE computer society press Los Alamitos}.
\newblock


\bibitem[McAllester(1998)]%
        {mcallester1998some}
\bibfield{author}{\bibinfo{person}{David~A McAllester}.}
  \bibinfo{year}{1998}\natexlab{}.
\newblock \showarticletitle{Some pac-bayesian theorems}. In
  \bibinfo{booktitle}{\emph{Proceedings of the eleventh annual conference on
  Computational learning theory}}. \bibinfo{pages}{230--234}.
\newblock


\bibitem[Meinke and Sindhu(2011)]%
        {meinke2011incremental}
\bibfield{author}{\bibinfo{person}{Karl Meinke} {and}
  \bibinfo{person}{Muddassar~A Sindhu}.} \bibinfo{year}{2011}\natexlab{}.
\newblock \showarticletitle{Incremental learning-based testing for reactive
  systems}. In \bibinfo{booktitle}{\emph{International Conference on Tests and
  Proofs}}. Springer, \bibinfo{pages}{134--151}.
\newblock


\bibitem[Miranda and Bertolino(2015)]%
        {miranda2015improving}
\bibfield{author}{\bibinfo{person}{Breno Miranda} {and}
  \bibinfo{person}{Antonia Bertolino}.} \bibinfo{year}{2015}\natexlab{}.
\newblock \showarticletitle{Improving test coverage measurement for reused
  software}. In \bibinfo{booktitle}{\emph{2015 41st Euromicro Conference on
  Software Engineering and Advanced Applications}}. IEEE,
  \bibinfo{pages}{27--34}.
\newblock


\bibitem[Miranda and Bertolino(2020)]%
        {miranda2020testing}
\bibfield{author}{\bibinfo{person}{Breno Miranda} {and}
  \bibinfo{person}{Antonia Bertolino}.} \bibinfo{year}{2020}\natexlab{}.
\newblock \showarticletitle{Testing relative to usage scope: Revisiting
  software coverage criteria}.
\newblock \bibinfo{journal}{\emph{ACM Transactions on Software Engineering and
  Methodology (TOSEM)}} \bibinfo{volume}{29}, \bibinfo{number}{3}
  (\bibinfo{year}{2020}), \bibinfo{pages}{1--24}.
\newblock


\bibitem[Mohri et~al\mbox{.}(2018)]%
        {mohri2018foundations}
\bibfield{author}{\bibinfo{person}{Mehryar Mohri}, \bibinfo{person}{Afshin
  Rostamizadeh}, {and} \bibinfo{person}{Ameet Talwalkar}.}
  \bibinfo{year}{2018}\natexlab{}.
\newblock \bibinfo{booktitle}{\emph{Foundations of machine learning}}.
\newblock \bibinfo{publisher}{MIT press}.
\newblock


\bibitem[Moore et~al\mbox{.}(1956)]%
        {moore1956gedanken}
\bibfield{author}{\bibinfo{person}{Edward~F Moore} {et~al\mbox{.}}}
  \bibinfo{year}{1956}\natexlab{}.
\newblock \showarticletitle{Gedanken-experiments on sequential machines}.
\newblock \bibinfo{journal}{\emph{Automata studies}}  \bibinfo{volume}{34}
  (\bibinfo{year}{1956}), \bibinfo{pages}{129--153}.
\newblock


\bibitem[Motwani and Raghavan(1996)]%
        {motwani1996randomized}
\bibfield{author}{\bibinfo{person}{Rajeev Motwani} {and}
  \bibinfo{person}{Prabhakar Raghavan}.} \bibinfo{year}{1996}\natexlab{}.
\newblock \showarticletitle{Randomized algorithms}.
\newblock \bibinfo{journal}{\emph{ACM Computing Surveys (CSUR)}}
  \bibinfo{volume}{28}, \bibinfo{number}{1} (\bibinfo{year}{1996}),
  \bibinfo{pages}{33--37}.
\newblock


\bibitem[Musa(1993)]%
        {musa1993operational}
\bibfield{author}{\bibinfo{person}{John~D. Musa}.}
  \bibinfo{year}{1993}\natexlab{}.
\newblock \showarticletitle{Operational profiles in software-reliability
  engineering}.
\newblock \bibinfo{journal}{\emph{IEEE software}} \bibinfo{volume}{10},
  \bibinfo{number}{2} (\bibinfo{year}{1993}), \bibinfo{pages}{14--32}.
\newblock


\bibitem[Nie and Leung(2011)]%
        {nie2011survey}
\bibfield{author}{\bibinfo{person}{Changhai Nie} {and} \bibinfo{person}{Hareton
  Leung}.} \bibinfo{year}{2011}\natexlab{}.
\newblock \showarticletitle{A survey of combinatorial testing}.
\newblock \bibinfo{journal}{\emph{ACM Computing Surveys (CSUR)}}
  \bibinfo{volume}{43}, \bibinfo{number}{2} (\bibinfo{year}{2011}),
  \bibinfo{pages}{1--29}.
\newblock


\bibitem[Reps et~al\mbox{.}(1997)]%
        {reps1997use}
\bibfield{author}{\bibinfo{person}{Thomas Reps}, \bibinfo{person}{Thomas Ball},
  \bibinfo{person}{Manuvir Das}, {and} \bibinfo{person}{James Larus}.}
  \bibinfo{year}{1997}\natexlab{}.
\newblock \showarticletitle{The use of program profiling for software
  maintenance with applications to the year 2000 problem}. In
  \bibinfo{booktitle}{\emph{Proceedings of the 6th European SOFTWARE
  ENGINEERING conference held jointly with the 5th ACM SIGSOFT international
  symposium on Foundations of software engineering}}.
  \bibinfo{pages}{432--449}.
\newblock


\bibitem[Romanik and Vitter(1993)]%
        {romanik1993using}
\bibfield{author}{\bibinfo{person}{Kathleen Romanik} {and}
  \bibinfo{person}{Jeffrey~Scott Vitter}.} \bibinfo{year}{1993}\natexlab{}.
\newblock \showarticletitle{Using computational learning theory to analyze the
  testing complexity of program segments}. In
  \bibinfo{booktitle}{\emph{Proceedings of 1993 IEEE 17th International
  Computer Software and Applications Conference COMPSAC'93}}. IEEE,
  \bibinfo{pages}{367--373}.
\newblock


\bibitem[Romanik and Vitter(1996)]%
        {romanik1996using}
\bibfield{author}{\bibinfo{person}{Kathleen Romanik} {and}
  \bibinfo{person}{Jeffrey~Scott Vitter}.} \bibinfo{year}{1996}\natexlab{}.
\newblock \showarticletitle{Using Vapnik--Chervonenkis Dimension to Analyze the
  Testing Complexity of Program Segments}.
\newblock \bibinfo{journal}{\emph{Information and Computation}}
  \bibinfo{volume}{128}, \bibinfo{number}{2} (\bibinfo{year}{1996}),
  \bibinfo{pages}{87--108}.
\newblock


\bibitem[Shalev-Shwartz and Ben-David(2014)]%
        {shalev2014understanding}
\bibfield{author}{\bibinfo{person}{Shai Shalev-Shwartz} {and}
  \bibinfo{person}{Shai Ben-David}.} \bibinfo{year}{2014}\natexlab{}.
\newblock \bibinfo{booktitle}{\emph{Understanding machine learning: From theory
  to algorithms}}.
\newblock \bibinfo{publisher}{Cambridge university press}.
\newblock


\bibitem[Shao et~al\mbox{.}(2022)]%
        {shao2022theory}
\bibfield{author}{\bibinfo{person}{Han Shao}, \bibinfo{person}{Omar Montasser},
  {and} \bibinfo{person}{Avrim Blum}.} \bibinfo{year}{2022}\natexlab{}.
\newblock \showarticletitle{A theory of pac learnability under transformation
  invariances}.
\newblock \bibinfo{journal}{\emph{Advances in Neural Information Processing
  Systems}}  \bibinfo{volume}{35} (\bibinfo{year}{2022}),
  \bibinfo{pages}{13989--14001}.
\newblock


\bibitem[Tramontana et~al\mbox{.}(2019)]%
        {tramontana2019developing}
\bibfield{author}{\bibinfo{person}{Porfirio Tramontana},
  \bibinfo{person}{Domenico Amalfitano}, \bibinfo{person}{Nicola Amatucci},
  \bibinfo{person}{Atif Memon}, {and} \bibinfo{person}{Anna~Rita Fasolino}.}
  \bibinfo{year}{2019}\natexlab{}.
\newblock \showarticletitle{Developing and evaluating objective termination
  criteria for random testing}.
\newblock \bibinfo{journal}{\emph{ACM Transactions on Software Engineering and
  Methodology (TOSEM)}} \bibinfo{volume}{28}, \bibinfo{number}{3}
  (\bibinfo{year}{2019}), \bibinfo{pages}{1--52}.
\newblock


\bibitem[Valiant(2014)]%
        {valiant2014probably}
\bibfield{author}{\bibinfo{person}{Leslie Valiant}.}
  \bibinfo{year}{2014}\natexlab{}.
\newblock \bibinfo{title}{Probably Approximately Correct: Nature's Algorithms
  for Learning and Prospering in a Complex World}.
\newblock
\newblock


\bibitem[Valiant(1984)]%
        {valiant1984theory}
\bibfield{author}{\bibinfo{person}{Leslie~G Valiant}.}
  \bibinfo{year}{1984}\natexlab{}.
\newblock \showarticletitle{A theory of the learnable}.
\newblock \bibinfo{journal}{\emph{Commun. ACM}} \bibinfo{volume}{27},
  \bibinfo{number}{11} (\bibinfo{year}{1984}), \bibinfo{pages}{1134--1142}.
\newblock


\bibitem[Vapnik(1999)]%
        {vapnik1999nature}
\bibfield{author}{\bibinfo{person}{Vladimir Vapnik}.}
  \bibinfo{year}{1999}\natexlab{}.
\newblock \bibinfo{booktitle}{\emph{The nature of statistical learning
  theory}}.
\newblock \bibinfo{publisher}{Springer science \& business media}.
\newblock


\bibitem[Walkinshaw and Fraser(2017)]%
        {walkinshaw2017uncertainty}
\bibfield{author}{\bibinfo{person}{Neil Walkinshaw} {and}
  \bibinfo{person}{Gordon Fraser}.} \bibinfo{year}{2017}\natexlab{}.
\newblock \showarticletitle{Uncertainty-driven black-box test data generation}.
  In \bibinfo{booktitle}{\emph{2017 IEEE International Conference on Software
  Testing, Verification and Validation (ICST)}}. IEEE,
  \bibinfo{pages}{253--263}.
\newblock


\bibitem[Weyuker(1983)]%
        {weyuker1983assessing}
\bibfield{author}{\bibinfo{person}{Elaine~J Weyuker}.}
  \bibinfo{year}{1983}\natexlab{}.
\newblock \showarticletitle{Assessing test data adequacy through program
  inference}.
\newblock \bibinfo{journal}{\emph{ACM Transactions on Programming Languages and
  Systems (TOPLAS)}} \bibinfo{volume}{5}, \bibinfo{number}{4}
  (\bibinfo{year}{1983}), \bibinfo{pages}{641--655}.
\newblock


\bibitem[Weyuker(1986)]%
        {weyuker1986axiomatizing}
\bibfield{author}{\bibinfo{person}{Elaine~J Weyuker}.}
  \bibinfo{year}{1986}\natexlab{}.
\newblock \showarticletitle{Axiomatizing software test data adequacy}.
\newblock \bibinfo{journal}{\emph{IEEE transactions on software engineering}}
  \bibinfo{number}{12} (\bibinfo{year}{1986}), \bibinfo{pages}{1128--1138}.
\newblock


\bibitem[Wong et~al\mbox{.}(2016)]%
        {wong2016survey}
\bibfield{author}{\bibinfo{person}{W~Eric Wong}, \bibinfo{person}{Ruizhi Gao},
  \bibinfo{person}{Yihao Li}, \bibinfo{person}{Rui Abreu}, {and}
  \bibinfo{person}{Franz Wotawa}.} \bibinfo{year}{2016}\natexlab{}.
\newblock \showarticletitle{A survey on software fault localization}.
\newblock \bibinfo{journal}{\emph{IEEE Transactions on Software Engineering}}
  \bibinfo{volume}{42}, \bibinfo{number}{8} (\bibinfo{year}{2016}),
  \bibinfo{pages}{707--740}.
\newblock


\bibitem[Wu et~al\mbox{.}(2022)]%
        {wu2022TCP}
\bibfield{author}{\bibinfo{person}{Penghao Wu}, \bibinfo{person}{Xiaosong Jia},
  \bibinfo{person}{Li Chen}, \bibinfo{person}{Junchi Yan},
  \bibinfo{person}{Hongyang Li}, {and} \bibinfo{person}{Yu Qiao}.}
  \bibinfo{year}{2022}\natexlab{}.
\newblock \bibinfo{title}{Trajectory-guided Control Prediction for End-to-end
  Autonomous Driving: A Simple yet Strong Baseline}.
\newblock
\newblock
\showeprint[arxiv]{2206.08129}~[cs.CV]


\bibitem[Zhu(1996)]%
        {zhu1996formal}
\bibfield{author}{\bibinfo{person}{Hong Zhu}.} \bibinfo{year}{1996}\natexlab{}.
\newblock \showarticletitle{A formal interpretation of software testing as
  inductive inference}.
\newblock \bibinfo{journal}{\emph{Software Testing, Verification and
  Reliability}} \bibinfo{volume}{6}, \bibinfo{number}{1}
  (\bibinfo{year}{1996}), \bibinfo{pages}{3--31}.
\newblock


\bibitem[Zhu et~al\mbox{.}(1992)]%
        {zhu1992inductive}
\bibfield{author}{\bibinfo{person}{Hong Zhu}, \bibinfo{person}{Patrick Hall},
  {and} \bibinfo{person}{John May}.} \bibinfo{year}{1992}\natexlab{}.
\newblock \showarticletitle{Inductive inference and software testing}.
\newblock \bibinfo{journal}{\emph{Software Testing, Verification and
  Reliability}} \bibinfo{volume}{2}, \bibinfo{number}{2}
  (\bibinfo{year}{1992}), \bibinfo{pages}{69--81}.
\newblock


\bibitem[Zhu et~al\mbox{.}(1997)]%
        {zhu1997software}
\bibfield{author}{\bibinfo{person}{Hong Zhu}, \bibinfo{person}{Patrick~AV
  Hall}, {and} \bibinfo{person}{John~HR May}.} \bibinfo{year}{1997}\natexlab{}.
\newblock \showarticletitle{Software unit test coverage and adequacy}.
\newblock \bibinfo{journal}{\emph{Acm computing surveys (csur)}}
  \bibinfo{volume}{29}, \bibinfo{number}{4} (\bibinfo{year}{1997}),
  \bibinfo{pages}{366--427}.
\newblock


\bibitem[Zhu et~al\mbox{.}(2022)]%
        {zhu2022fuzzing}
\bibfield{author}{\bibinfo{person}{Xiaogang Zhu}, \bibinfo{person}{Sheng Wen},
  \bibinfo{person}{Seyit Camtepe}, {and} \bibinfo{person}{Yang Xiang}.}
  \bibinfo{year}{2022}\natexlab{}.
\newblock \showarticletitle{Fuzzing: a survey for roadmap}.
\newblock \bibinfo{journal}{\emph{ACM Computing Surveys (CSUR)}}
  \bibinfo{volume}{54}, \bibinfo{number}{11s} (\bibinfo{year}{2022}),
  \bibinfo{pages}{1--36}.
\newblock


\end{thebibliography}

\end{document}